\documentclass[11pt, journal,draftcls,onecolumn]{IEEEtran}

\usepackage{cite}
\ifCLASSINFOpdf
\else
\fi

\usepackage{array}

\ifCLASSOPTIONcompsoc
  \usepackage[caption=false,font=normalsize,labelfont=sf,textfont=sf]{subfig}
\else
  \usepackage[caption=false,font=footnotesize]{subfig}
\fi

\usepackage[hidelinks]{hyperref}
\usepackage{url}

\usepackage{graphicx} 
\graphicspath{{./Figures/}}
\usepackage{mathtools,amsmath,amssymb,amsthm} 
\usepackage[overload]{empheq}
\usepackage{enumitem}
\usepackage{multicol}
\usepackage{capt-of}
\usepackage[export]{adjustbox}

\newtheorem{proposition}{Proposition}[section]
\newtheorem{definition}{Definition}[section]
\theoremstyle{remark}
\newtheorem*{remark}{Remark}

\begin{document}
%
\title{High-Order Synchrosqueezing Transform for Multicomponent Signals Analysis - With an Application to Gravitational-Wave Signal}
%
%
%
\author{Duong-Hung~Pham, and
        Sylvain~Meignen. 
\thanks{~D-H Pham and S. Meignen are with the Jean Kuntzmann Laboratory, University
of Grenoble-Alpes, and CNRS, Grenoble 38041, France (\href{mailto:email:duong-hung.pham@imag.fr}{email:duong-hung.pham@imag.fr} and \href{mailto:sylvain.meignen@imag.fr}{sylvain.meignen@imag.fr}). The authors acknowledge the support of the French Agence Nationale de la Recherche (ANR) 
under reference ANR-13- BS03-0002-01 (ASTRES)}
}

\markboth{IEEE Transactions on Signal Processing}%
{Shell \MakeLowercase{\textit{et al.}}: Bare Demo of IEEEtran.cls for IEEE Journals}
%


\maketitle

\begin{abstract}
This study puts forward a generalization of the short-time Fourier-based Synchrosqueezing Transform using a new local estimate of instantaneous frequency. Such a technique enables not only to achieve a highly concentrated time-frequency representation for a wide variety of AM-FM multicomponent signals but also to reconstruct their modes with a high accuracy. Numerical investigation on synthetic and gravitational-wave signals shows the efficiency of this new approach.   
\end{abstract}


\IEEEpeerreviewmaketitle

\section{Introduction}
\IEEEPARstart{M}{any} signals such as audio signals (music, speech), medical data (electrocardiogram, thoracic and abdominal movement signals), can be modeled as a superposition of amplitude- and frequency-modulated (AM-FM) modes \cite{Meignen2012a,Lin2016, Herry2017}, called multicomponent signals (MCS). Linear techniques as for instance continuous wavelet transforms (CWT) and short-time Fourier transform (STFT) are often utilized to characterize such signals in the time-frequency (TF) plane. However, they all share the same limitation, known as the ``uncertainty principle", stipulating that one cannot localize a signal with arbitrary precision both in time and frequency. Many efforts were made to cope with this issue and, in particular, a general methodology to sharpen TF representation, called ``reassignment" method (RM) was proposed. This was first introduced in \cite{Kodera1978}, in a somehow restricted framework, and then further developed in \cite{Auger1995}, as a post-processing technique. The main problem associated with RM is that the reassigned transform is no longer invertible and does not allow for mode reconstruction.     

In the context of audio signal analysis \cite{Daubechies1996}, Daubechies and Maes proposed another phase-based technique, called ``SynchroSqueezing Transform" (SST), whose theoretical analysis followed in \cite{Daubechies2011}. Its purpose is relatively similar to that of RM, i.e. to sharpen the time-scale (TS) representation given by CWT, with the additional advantage of allowing for mode retrieval. Using the principle of wavelet-based SST (WSST), Thakur and Wu proposed an extension of SST to the TF representation given by STFT (FSST) \cite{Thakur2011}, which was then proven to be robust to small bounded perturbations and noise \cite{Thakur2013}. Nevertheless, the applicability of SST is somewhat hindered by the requirement of weak frequency modulation hypothesis for the modes constituting the signal. In contrast, most real signals are made up of very strongly modulated AM-FM modes, as for instance chirps involved in radar \cite{Skolnik2008}, speech processing \cite{Pitton1994}, or gravitational waves \cite{Candes2008,Abbott2016}. In this regard, a recent adaptation of FSST to the context of strongly modulated modes was introduced in \cite{Oberlin2015}, and further mathematically analyzed in \cite{Behera2015}. Unfortunately, the aforementioned technique was proven to only provide an ideal invertible TF representation for linear chirps with Gaussian modulated amplitudes, which is still restrictive.                 

In this paper, we propose to improve existing STFT-based SSTs by computing more accurate estimates of the instantaneous frequencies of the modes making up the signal, using higher order approximations both for the amplitude and phase. This results in perfect concentration and reconstruction for a wider variety of AM-FM modes than 
what was possible up to now with synchrosqueezing techniques. 

This paper is structured as follows: we recall some fundamental notation and definitions on Fourier Transform (FT), STFT and MCS in Section \ref{sec:definitions}, and  introduce FSST with its extension, the second-order FSST (FSST2) respectively in Sections \ref{sec:STFT-based_SST} and \ref{sec:FSST2}. We then present the proposed generalization, called {\it higher-order synchrosqueezing transform} in Section \ref{sec:higher_order_SST}. Finally, the numerical simulations of Section \ref{sec:numerical_implementation} demonstrate the interest of our technique on both simulated signals and a gravitational-wave signal. 

\section{Background to FSST}
Before going in detail into the principle of FSST, the following section presents several notation that will be used in the sequel. 

\subsection{Basic Notation and Definitions} \label{sec:definitions}
The Fourier transform (FT) of a given signal $f \in L^1({\mathbb{R})}$ is defined as: 
\begin{equation} 
\hat{f}(\eta) = \int_\mathbb{R}{f(t)} e^{-i2\pi \eta t}dt.
\end{equation}
If $\hat{f}$ is also integrable, $f$ can be reconstructed through: 
\begin{equation}
f(t) =  \int_\mathbb{R} \hat f (\eta) e^{i2\pi \eta t}d\eta. 
\end{equation}
It is well known that time- or frequency-domain representation alone is not appropriate to describe non-stationary signals whose frequencies have a temporal localization.  The short-time Fourier transform (STFT) was thus introduced for that purpose, and is defined as follows: given a signal $f \in L^1(\mathbb{R})$ and a window $g$ in the Schwartz class, the space of smooth functions with  fast decaying derivatives of any order, the (modified) STFT of $f$ is defined by:
\begin{equation} \label{eqn:STFT}
V^g_f(t,\eta) = \int_\mathbb{R} f(\tau)g^*(\tau-t)e^{-2i\pi \eta (\tau-t)}d\tau,
\end{equation}
where $g^*$ is the complex conjugate of $g$, and then the spectrogram corresponds to $|V_f^g(t,\eta)|^2$. Furthermore, the original signal $f$ can be retrieved from its STFT through the following synthesis formula, on condition that $g$ does not vanish and is continuous at $0$:
\begin{equation} \label{eqn:synthesis_STFT}
 f(t) = \frac{1}{g^*(0)} \int_\mathbb{R} V^g_f(t,\eta) d\eta.
\end{equation}
If $f$ is analytic, i.e. $\eta \le 0$ then $\hat{f}(\eta)=0$, the integral in (\ref{eqn:synthesis_STFT}) only takes place on $\mathbb{R}_+$. 

In the sequel, we will intensively study multicomponent signals (MCS) defined as a superposition of AM-FM components or modes:
 \begin{equation} 
 f(t) = \sum \limits_{k=1}^K f_k(t)~~~~\textit{with}~~f_k(t) = A_k(t)e^{i 2\pi \phi_k(t)},
\end{equation}
for some finite $K\in \mathbb{N}$, $A_k(t)$ and $\phi_k(t)$ are respectively instantaneous amplitude (IA) and phase (IP) functions satisfying: $A_k(t)>0,\phi'_k(t)>0$ and $\phi'_{k+1}(t)>\phi'_k(t)$ for all $t$ where $\phi'_k(t)$ is referred to as the instantaneous frequency (IF) of mode $f_k$ at time $t$. Such a signal is fully described by its ideal TF (ITF) representation defined as:
\begin{align}
 \mathrm{TI}_f(t,\omega) = \sum \limits_{k=1}^K A_k(t) \delta \left(\omega-\phi'_k(t)\right),
\end{align}
where $\delta$ denotes the Dirac distribution.   

\subsection{STFT-based SST (FSST)} \label{sec:STFT-based_SST}
The key idea of STFT-based SST (FSST) is to sharpen the ``blurred'' STFT representation of $f$ by using the following IF estimate at time $t$ and frequency $\eta$:
\begin{equation}
 \hat{\omega}_f(t,\eta) = \frac{1}{2\pi} \partial_t\text{arg}\left \{ V_f^{g}(t,\eta) \right \} = \Re \left\{\frac{\partial_t V_f^g(t,\eta)}{2 i \pi V_f^g(t,\eta)} \right\},
\end{equation}
where $\text{arg}\{ Z\}$ and $\Re\{Z\}$ stand for the argument and real part of complex number $Z$, respectively, and $\partial_t$ is the partial derivative with respect to $t$. 

Indeed, $V^g_f(t,\eta)$ is reassigned to a new position  $(t,\hat{\omega}_f(t,\eta))$ using the synchrosqueezing operator defined as follows:
\begin{align} 
\label{eqn:SST_operator}
 &T_{f}^{g,\gamma}(t,\omega) =  \dfrac{1}{g^*(0)} \int_{\{\eta,|V_f^g(t,\eta)|>\gamma\}} V_f^g(t,\eta) \delta \left(\omega-\hat{\omega}_f(t,\eta)\right)d \eta,
\end{align}
where $\gamma$ is some threshold. 

Since its coefficients are reassigned along the ``frequency'' axis, FSST preserves the causality property, thus making the $k^{th}$ mode approximately reconstructed by integrating $T_{f}^{g,\gamma}(t,\eta)$ in the vicinity of the corresponding ridge $(t,\phi_k'(t))$:
\begin{equation} 
\label{eqn:reconstruction_formula_SST}
 f_k(t) \approx \int_{\{\omega,|\omega-\varphi_k(t)|<d\}}T_{f}^{g,\gamma}(t,\omega)d\omega,
\end{equation}
where $\varphi_k(t)$ is an estimate of $\phi_k'(t)$. Parameter $d$ enables to compensate for both the inaccurate approximation $\varphi_k(t)$ of $\phi_k'(t)$ and the error made by estimating the IF by means of $\hat{\omega}_f(t,\eta)$. It is worth noting here that the approximation $\varphi_k(t)$ must be computed before retrieving mode $f_k$. For that purpose, a commonly used technique is based on ridge extraction assuming $T_{f}^{g,\gamma}$ and $K$ are known \cite{Daubechies2011,Auger2013}. This technique initially proposed by Carmona et {\itshape al.} \cite{Carmona1997} relies on the minimization of the following energy functional:
\begin{align} \label{eqn:energy_functional}
 E_f(\varphi) = \sum \limits_{k=1}^K - \int_\mathbb{R}|T_{f}^{g,\gamma}(t,\varphi_k(t))|^2dt+\int_\mathbb{R}\lambda \varphi_k'(t)^2+\beta \varphi_k''(t)^2dt,
\end{align}
where $\lambda$ and $\beta$ are chosen regularization parameters such that the trade-off between smoothness of $\varphi_k$ and energy is maximized. In practice, this energy functional is hard to implement because of its non-convexity, and so one should find tricks to avoid local minima as much as possible, as for example using a  simulated annealing algorithm proposed in \cite{Carmona1997}. A recent algorithm introduced in \cite{Thakur2013}, and used in this paper, determines the ridge associated with the corresponding mode thanks to a forward/backward approach for different initializations. Furthermore, a detailed study on the influence of the regularization parameters, introduced recently in \cite{Meignen2016}, shows that they should not be used in (\ref{eqn:energy_functional}) since they bring no improvement in term of accuracy of ridge estimation. 

Finally, a very important aspect of FSST is that it is developed in a solid mathematical framework. Indeed, assume the modes of an MCS satisfy the following definition: 
\begin{definition} \label{eqn:condition_SST}
Let $\epsilon>0$ and $\mathcal{B}_{\varepsilon,\Delta}$ be the class of MCS such that for all $k$, $A_k\in C^{1}(\mathbb{R})\bigcap L^{\infty}(\mathbb{R}), ~\phi_k \in C^{2}(\mathbb{R}),~\textit{supp}_t \phi'_k(t) < \infty,~\textit{and~ for}~\forall t, ~ A_k(t)>0, ~\phi'_k(t)>0$ satisfy two hypotheses: 
\setlist[itemize]{leftmargin=*}
 \begin{itemize}
  \item H1) $f_k$s have weak frequency modulation, i.e. $\exists \varepsilon $ small s.t.: $|A_k'(t)| \le \varepsilon~\textit{and}~|\phi_k''(t)| \le \varepsilon~\textit{for}~\forall t. $
  \item H2) all $f_k$s are well separated in frequency, i.e. $ |\phi_{k+1}'(t)-\phi_k'(t)| \ge 2\Delta~\textit{for}~\forall t ~\textit{and}~ \forall k \in \{1,...,K\}$, where $\Delta$ is called the separation parameter. 
 \end{itemize}
\end{definition}
Then, it was proven in \cite{Behera2015} that the synchrosqueezing operator $T_{f}^{g,\gamma}$ is concentrated in narrow bands around the curves $(t,\phi_k'(t))$ in the TF plane and the modes $f_k$s can be reconstructed from $T_{f}^{g,\gamma}(t,\omega)$ with a reasonably high accuracy. 

\subsection{Second Order STFT-based SST (FSST2)} \label{sec:FSST2}
Although FSST proves to be an efficient solution for enhancing TF representations, its application is restricted to a class of MCS composed of slightly perturbed pure harmonic modes. To overcome this limitation, a recent extension of FSST was introduced based on a more accurate IF estimate, which is then used to define an improved synchrosqueezing operator $T_{2,f}^{g,\gamma}(t,\eta)$, called second-order STFT-based synchrosqueezing transform (FSST2) \cite{Oberlin2015, Behera2015}. 

More precisely, a second-order local modulation operator is first defined and then used to compute the new IF estimate. This modulation operator corresponds to the ratio of the 
first-order derivatives, with respect to $t$, of the reassignment operators, as explained in the following: 

\begin{proposition} Given a signal $f \in L^2(\mathbb{R})$, the complex reassignment operators $\tilde{\omega}_f(t,\eta)$ and $\tilde{\tau}_f(t,\eta)$ are respectively defined for any $(t,\eta)$ s.t. $V^{g}_{f}(t,\eta) \ne 0$ as:
\begin{align} \label{eqn:complex_IF_nSST_general_definition}
\tilde{\omega}_f(t,\eta) &=  \frac{\partial_t V_f^g(t,\eta)}{2 i \pi V_f^g(t,\eta)} \notag\\
\tilde{\tau}_f(t,\eta) & = t - \frac{\partial_\eta V_f^g(t,\eta)}{2 i \pi V_f^g(t,\eta)}.
\end{align}
Then, the second-order local complex modulation operator $\tilde{q}_{t,f}(t,\eta)$ is defined by:
\begin{align} \label{eqn:FSST2_local_operator}
\tilde{q}_{t,f}(t,\eta) = \frac{\partial_t \tilde \omega_f (t,\eta)}{\partial_t \tilde{\tau}_f(t,\eta)} ~~~~~\text{whenever}~\partial_t \tilde{\tau}_f(t,\eta) \ne 0.
\end{align}
\end{proposition}

In that case, the definition of the improved IF estimate associated with the TF representation given by STFT is derived as: 
\begin{definition}
Let $f \in L^2(\mathbb{R})$, the second-order local complex IF estimate of  f is defined as: 
\begin{equation}[left = {\tilde{\omega}_{t,f}^{[2]}(t,\eta) =} \empheqlbrace ]
  \begin{alignedat}{2}
      &  \tilde \omega_f(t,\eta) + \tilde{q}_{t,f}(t,\eta)(t-\tilde{\tau}_f(t,\eta)) & ~& \text{if}   ~\partial_t\tilde{\tau}_f \ne 0\\
      & \tilde{\omega}_f(t,\eta)  &  &    \text{otherwise.}
  \end{alignedat} \notag
\end{equation}
Then, its real part $\hat{\omega}_{t,f}^{[2]}(t,\eta)=\Re\{\tilde{\omega}_{t,f}^{[2]}(t,\eta)\}$ is the desired IF estimate.  
\end{definition}
It was demonstrated in \cite{Oberlin2015} that $\Re \left\{ \tilde{q}_{t,f}(t,\eta) \right\} = \phi''(t)$ when $f$ is a Gaussian modulated linear chirp, i.e. $f(t) = A(t)e^{i 2\pi \phi(t)}$ where both $\log (A(t))$ and $\phi(t)$ are quadratic. Also,  $\Re\{\tilde{\omega}_{t,f}^{[2]}(t,\eta)\}$ is an exact estimate of $\phi'(t)$ for this kind of signals. For a more general mode with Gaussian amplitude, its IF can be estimated by $\Re\{\tilde{\omega}_{t,f}^{[2]}(t,\eta)\}$, in which the estimation error only involves the derivatives of the phase with orders larger than 3. Furthermore, $\tilde{\omega}_f$, $\tilde{\tau}_f$ and $\tilde{q}_{t,f}$ can be computed by means of only five STFTs as follows:
\begin{proposition} \label{propo:complex_IF_nSST_efficient_definition}
For a signal $f \in L^2(\mathbb{R})$, the expressions $\tilde{\omega}_f$, $\tilde{\tau}_f$ and $\tilde{q}_{t,f}$ can be written as: 
\begin{align} 
 \tilde{\omega}_f  &=\eta - \frac{1}{i2\pi} \frac{V_f^{g'}}{V_f^g}\\
 \tilde{\tau}_f  &=t + \frac{V_f^{tg}}{V_f^g}\\
 \tilde{q}_{t,f} &= \frac{1}{i2\pi}\frac{V_f^{g''}V_f^{g}-\left(V_f^{g'}\right)^2}{V_f^{tg}V_f^{g'}-V_f^{tg'}V_f^{g}},
\end{align} 
where $V_f^{g}$ denotes $V_f^{g}(t,\eta)$ and $V_f^{g'}, V_f^{tg}, V_f^{g''}, V_f^{tg'}$ are respectively STFTs of $f$ computed with windows $t \mapsto g'(t), tg(t), g''(t)$ and $tg'(t)$. 
\end{proposition}

The second-order FSST (FSST2) is then defined by simply replacing $\hat{\omega}_f(t,\eta)$  by $\hat{\omega}_{t,f}^{[2]}(t,\eta)$ in (\ref{eqn:SST_operator}):
\begin{align} \label{eqn:FSST2}
 &T_{2, f}^{g,\gamma}(t,\omega) =  \dfrac{1}{g^*(0)} \int_{\{\eta,|V_f^g(t,\eta)|>\gamma\}} V_f^g(t,\eta) \delta \left(\omega-\hat{\omega}_{t,f}^{[2]}(t,\eta)\right) d\eta. \notag
\end{align}

Mode $f_k$ is finally retrieved by replacing $T_{f}^{g,\gamma}(t,\omega)$ by $T_{2, f}^{g,\gamma}(t,\omega)$ in (\ref{eqn:reconstruction_formula_SST}). Note that the theoretical foundation to support FSST2 has just been proposed in \cite{Behera2015}.

\begin{remark} By using partial derivatives with respect to $\eta$ instead of $t$, a new second-order local modulation operator $\tilde{q}_{\eta,f}(t,\eta)$ showing the same properties as those of $\tilde{q}_{t,f}(t,\eta)$ can be obtained as follows:
\begin{definition} \label{def:FSST2_local_operator_frequency} Given a signal $f \in L^2(\mathbb{R})$, the second-order local complex modulation operator $\tilde{q}_{\eta,f}$ is defined by:
\begin{align} \label{eqn:FSST2_local_operator_frequency}
\tilde{q}_{\eta,f} (t,\eta) &=\dfrac{\partial_{\eta} \tilde{\omega}_f (t,\eta)}{\partial_{\eta} \tilde{\tau}_f (t,\eta)}~~~~~\text{whenever}~\partial_{\eta} \tilde{\tau}_f(t,\eta) \ne 0,
\end{align}
where $\tilde{\omega}_f(t,\eta)$ and $\tilde{\tau}_f(t,\eta)$ are respectively defined in (\ref{eqn:complex_IF_nSST_general_definition}). 
\end{definition}

The next proposition shows that this new operator also leads to a perfect estimate of the frequency modulation for a Gaussian modulated linear chirp.

\begin{proposition}  If $f(t) = A(t)e^{i 2\pi \phi(t)}$ is a Gaussian modulated linear chirp, then $ \Re \left\{ \tilde{q}_{\eta,f} (t,\eta) \right\} = \phi''(t)$.      
\end{proposition}
\begin{IEEEproof}  Let us consider a mode $f(\tau) =A(\tau) e^{i2\pi\phi(\tau)}$ where $\log (A(\tau))$ and $\phi(\tau)$ are quadratic functions described by: 
\begin{align}
 \log(A(\tau)) = \sum \limits_{k=0}^{2} \dfrac{\alpha_k}{k!} \tau^k~~~\text{and}~~~ \phi(\tau) = \sum \limits_{k=0}^{2} \dfrac{\beta_k}{k!} \tau^k,  \notag
 \end{align}
with $\alpha_k, \beta_k \in \mathbb{R}$. The STFT of this mode with any window $g$, at time $t$ and frequency $\eta$, can be written as:  
\begin{align}
& V_{f}^g(t,\eta) = \int_\mathbb{R} f(\tau+t) g(\tau) e^{-i2\pi\eta\tau} d\tau\notag\\
&= \int_\mathbb{R}  \exp \left(\sum \limits_{k=0}^{2} \dfrac{1}{k!} \left(\alpha_k + i2\pi \beta_k \right)(\tau+t)^k \right) g(\tau) e^{-i2\pi\eta \tau} d\tau. \notag
 \end{align}
By taking the partial derivative of $V_{f}^g(t,\eta)$ with respect $t$, and then dividing by $i2\pi V_{f}^g(t,\eta)$, the local IF estimate  $\tilde{\omega}_{f}(t,\eta)$ defined in (\ref{eqn:complex_IF_nSST_general_definition}) can be obtained for $V^{g}_{f}(t,\eta) \ne 0$:
 \begin{align}  \label{eqn:complex_IF_STFT}
\tilde{\omega}_f (t,\eta) &= \sum \limits_{k=1}^{2} \left(\frac{1}{i2\pi}\alpha_k + \beta_k \right) t^{k-1} + \left(\frac{1}{i2\pi}\alpha_2 + \beta_2 \right)  \frac{V_{f}^{tg}(t,\eta)}{V_{f}^g(t,\eta)}.
\end{align} 
Then, taking the partial derivative of (\ref{eqn:complex_IF_STFT}) with respect to $\eta$ and recalling from Proposition \ref{propo:complex_IF_nSST_efficient_definition} that $ \dfrac{V_f^{tg}(t,\eta)}{V_f^g(t,\eta)} = \tilde{\tau}_f(t,\eta)  - t$, we get the following expression:
 \begin{align} \label{eqn:complex_IF_STFT_1} 
 \partial_{\eta} \tilde{\omega}_f (t,\eta) &= \left(\frac{1}{i2\pi}\alpha_2 + \beta_2 \right)\partial_{\eta}  \tilde{\tau}_f(t,\eta).  
\end{align} 
Setting $\tilde{q}_{\eta,f}(t,\eta) =\dfrac{\partial_{\eta} \tilde{\omega}_f(t,\eta)}{\partial_{\eta} \tilde{\tau}_f (t,\eta)}$ assuming $\partial_{\eta} \tilde{\tau}_f (t,\eta) \ne 0$ and noting that $\beta_2 = \phi''(t)$, ends the proof.  
\end{IEEEproof}

From (\ref{eqn:complex_IF_STFT}) and (\ref{eqn:complex_IF_STFT_1}), we also have the following result: 
\begin{align} 
\phi'(t) &= \beta_1 +\beta_2 t  \notag\\
&= \Re \left\{ \tilde{\omega}_f (t,\eta) - \left(\frac{1}{i2\pi}\alpha_2 + \beta_2 \right) (\tilde{\tau}_f (t,\eta) - t) \right\}\notag\\
&=  \Re \left\{ \tilde{\omega}_f (t,\eta) + \tilde{q}_{\eta,f}(t,\eta)(t-\tilde{\tau}_f (t,\eta)) \right\}.
\end{align}
Putting $\tilde{\omega}_{\eta,f}^{[2]}(t,\eta) = \tilde{\omega}_f (t,\eta) + \tilde{q}_{\eta,f}(t,\eta)(t-\tilde{\tau}_f (t,\eta))$, it follows that $\phi'(t) = \Re \left\{ \tilde{\omega}_{\eta,f}^{[2]}(t,\eta) \right\}$. Thus, a new IF estimate having the same properties as $\tilde{\omega}_{t,f}^{[2]}(t,\eta)$ is introduced as follows: 
\begin{definition} \label{def:FSST2_local_IF_estimate}
Let $f \in L^2(\mathbb{R})$, the second-order local complex IF estimate of signal f is defined by: 
\begin{equation}[left = {\tilde{\omega}_{\eta,f}^{[2]}(t,\eta) =} \empheqlbrace ]
  \begin{alignedat}{2}
      &  \tilde \omega_f(t,\eta) + \tilde{q}_{\eta,f}(t,\eta)(t-\tilde{\tau}_f(t,\eta)) & ~& \text{if}   ~\partial_{\eta}\tilde{\tau}_f(t,\eta) \ne 0\\
      & \tilde{\omega}_f(t,\eta)  &  &    \text{otherwise.}
  \end{alignedat} \notag
\end{equation}
Then, its real part $\hat{\omega}_{\eta,f}^{[2]}(t,\eta)=\Re\{\tilde{\omega}_{\eta,f}^{[2]}(t,\eta)\}$ is the desired IF estimate.  
\end{definition}

Note, finally, that: 
\begin{proposition} \label{propo:FSST2_local_operator_frequency} The second-order modulation operator $\tilde{q}_{\eta,f}(t,\eta)$ can be computed by:   
\begin{align} 
\tilde{q}_{\eta,f} = \dfrac{1}{i2\pi}\dfrac{\left(V_{f}^g\right)^2 + V_{f}^{g}V_{f}^{tg'}-V_{f}^{g'}V_{f}^{tg}}{V_{f}^{g}V_{f}^{t^2g}-\left(V_{f}^{tg}\right)^2}, 
\end{align} 
where $V_f^{t^2g}$ is the STFT of the signal $f$ computed with window $t \mapsto t^2g(t)$. 
\end{proposition}
\begin{IEEEproof}
By computing the partial derivatives of $\tilde{\omega}_f(t,\eta)$  and $ \tilde{\tau}_f(t,\eta)$ with respect to $\eta$ in the expressions given in Proposition \ref{propo:complex_IF_nSST_efficient_definition}, and then using formula $
\partial_{\eta} V_{f}^{g}(t,\eta) = -i2\pi V_{f}^{tg}(t,\eta)$, the expression for $\tilde{q}_{\eta,f}$ follows.  
\end{IEEEproof}

\end{remark}

\section{Higher Order Synchrosqueezing Transform}\label{sec:higher_order_SST}
Despite FSST2 definitely improves the concentration of TF representation, it is only demonstrated to work well on perturbed linear chirps with Gaussian modulated amplitudes. To handle signals containing more general types of AM-FM modes having non-negligible  $\phi_k^{(n)}(t)$ for $n \geq 3$, we are going to define new synchrosqueezing operators,  based on approximation orders higher than three for both amplitude and phase.

\subsection{Nth-order IF Estimate} 
The new IF estimate  we define here is based on high order Taylor expansions of the amplitude and phase of a mode. For that purpose, let us first consider a mode defined as in the following:   
\begin{definition} \label{def:modeled_signal_FSSTN} Given a mode $f(\tau) =A(\tau) e^{i2\pi\phi(\tau)}$ in $L^2(\mathbb{R})$ with $A(\tau)$ (\textit{resp}. $\phi(\tau)$) equal to its $L^{th}$-order (\textit{resp}. $N^{th}$-order) Taylor expansion for $\tau$ close to $t$:  
\begin{align}
\log (A(\tau)) &= \sum \limits_{k=0}^{L} \frac{[\log(A)]^{(k)}(t)}{k!}\left(\tau-t\right)^k \notag\\
\phi(\tau) &= \sum \limits_{k=0}^{N} \frac{\phi^{(k)}(t)}{k!}\left(\tau-t\right)^k \notag
\end{align}
where $Z^{(k)}(t)$ denotes the $k^{th}$ derivative of $Z$ evaluated at $t$.  
\end{definition}

A mode $f$ defined as above, with  $L \leq N$, can be written as: 
\begin{align}
f(\tau) &= \exp \left(\sum \limits_{k=0}^{N} \frac{1}{k!}\left( [\log(A)]^{(k)}(t) + i2\pi \phi^{(k)}(t) \right) \left(\tau-t\right)^k \right), \notag
\end{align}
since $[\log(A)]^{(k)}(t) = 0$ if $ L+1 \le k \le N$. 
Consequently, the STFT of this mode at time $t$ and frequency $\eta$ can be written as: 
\begin{align}
V_{f}^g(t,\eta)& =  \int_\mathbb{R} f(\tau+t) g(\tau) e^{-i2\pi\eta\tau} d\tau\notag\\
&= \int_\mathbb{R}  \exp \left(\sum \limits_{k=0}^{N} \frac{1}{k!}\left( [\log(A)]^{(k)}(t) + i2\pi \phi^{(k)}(t) \right) \tau^k \right)
g(\tau) e^{-i2\pi\eta \tau} d\tau. \notag
 \end{align}
By taking the partial derivative of $V_{f}^g(t,\eta)$ with respect to $t$ and then dividing by $i2\pi V_{f}^g(t,\eta)$, the local IF estimate $\tilde{\omega}_{f}(t,\eta)$ defined in (\ref{eqn:complex_IF_nSST_general_definition}) can be written when $V^{g}_{f}(t,\eta) \ne 0$ as:
 \begin{align}  \label{eqn:complex_IF_nSST_general}
 \tilde{\omega}_{f} (t,\eta) &= \sum \limits_{k=1}^{N} r_k(t) \frac{V_{f}^{t^{k-1} g}(t,\eta)}{V_{f}^{g}(t,\eta)}\notag\\
 &= \frac{1}{i2\pi} [\log(A)]'(t) + \phi'(t) +\sum \limits_{k=2}^{N} r_k(t) \frac{V_{f}^{t^{k-1} g}(t,\eta)}{V_{f}^{g}(t,\eta)},
\end{align}
where $r_{k}(t)$ are functions of $t$ defined for $k = 1, \hdots, N$ as: 
\begin{align} 
r_{k}(t)= \frac{1}{(k-1)!}\left( \dfrac{1}{i2\pi} [\log(A)]^{(k)}(t)  + \phi^{(k)}(t) \right).\notag 
\end{align}

It is clear from (\ref{eqn:complex_IF_nSST_general}) that, since $A(t)$ and $\phi(t)$ are real expressions, $\Re \left\{\tilde{\omega}_f (t,\eta)\right\} = \phi'(t)$ does not hold when the sum on the right hand side of (\ref{eqn:complex_IF_nSST_general}) has a non-zero real part. As in the case of the Gaussian modulated linear chirp introduced before, to get the exact IF estimate for the studied signal, one needs to subtract $\Re \left\{\sum \limits_{k=2}^{N} r_k(t) \dfrac{V_{f}^{t^{k-1} g}(t,\eta)}{V_{f}^{g}(t,\eta)} \right\}$ to 
$\Re \left\{\tilde{\omega}_f (t,\eta) \right\}$, for which $r_k(t)$, for all $k = 2, \hdots, N$, must be estimated.  

For that purpose, inspired by our study of the Gaussian modulated linear chirp, we derive a frequency modulation operator $\tilde{q}_{\eta,f}^{[k,N]}$, equal to $r_k(t)$ when $f$  satisfies Definition \ref{def:modeled_signal_FSSTN}, obtained by differentiating different STFTs with respect to $\eta$, as explained hereafter. Note that we choose to differentiate  
with respect to $\eta$ rather than $t$ because it leads to much simpler expressions, mainly as a result of the following formulae:   
\begin{align} 
\partial_t V_{f}^g(t,\eta) &= i2\pi \eta V_{f}^g(t,\eta) - V_{f}^{g'}(t,\eta)\notag\\
\partial_{\eta} V_{f}^{g}(t,\eta)& = -i2\pi V_{f}^{tg}(t,\eta).
\end{align}

The different modulation operators $\tilde{q}_{\eta,f}^{[k,N]}$ for $k = 2, \hdots, N$ can then be derived recursively, as explained in the next proposition: 
\begin{proposition} Given a mode $f \in L^2(\mathbb{R})$ that satisfies Definition \ref{def:modeled_signal_FSSTN} with $L \le N$, the $N-1$ local modulation operators $\tilde{q}_{\eta,f}^{[k,N]}$ such that $\Re \left\{ \tilde{q}_{\eta,f}^{[k,N]}(t,\eta) \right\} = \dfrac{\phi^{(k)}(t)}{(k-1)!}$, $k = 2, \hdots, N$, can be determined by:  \label{propo:estimates_pj}
\begin{align} 
\tilde{q}_{\eta,f}^{[N,N]}(t,\eta) &= y_{N}(t,\eta)~\text{and} \notag\\
\tilde{q}_{\eta,f}^{[j,N]}(t,\eta) &= y_{j}(t,\eta) - \sum \limits_{k=j+1}^N x_{k,j}(t,\eta) \tilde{q}_{\eta,f}^{[k,N]}(t,\eta) ~~~~~~~\text{for}~j = N-1, N-2, \hdots, 2,~\notag
\end{align}
where $y_{j}(t,\eta)$ and $x_{k,j}(t,\eta)$ are defined as follows. For any $(t,\eta)$ s.t. $V^{g}_{f}(t,\eta) \ne 0$ and $\partial_{\eta}x_{j,j-1}(t,\eta) \ne 0$, we put:    
\begin{align}
&\text{for}~ k=1 \hdots N,~~ y_{1}(t,\eta)  = \tilde{\omega}_{f}(t,\eta) ~ \text{and} ~ x_{k,1}(t,\eta) = \frac{V_{f}^{t^{k-1} g}(t,\eta)}{V_{f}^{g}(t,\eta)}, \notag\\
&\text{and then}~\text{for}~ j=2 \hdots N ~ \text{and} ~ k=j \hdots N,\notag\\
&y_{j}(t,\eta)  = \dfrac{\partial_{\eta} y_{j-1}(t,\eta) }{\partial_{\eta}x_{j,j-1}(t,\eta)} ~ \text{and} ~ x_{k,j}(t,\eta) = \dfrac{\partial_{\eta}x_{k,j-1}(t,\eta)}{\partial_{\eta}x_{j,j-1}(t,\eta)}. ~ \notag  
\end{align}
\end{proposition}

The proof of Proposition \ref{propo:estimates_pj} is given in Appendix \ref{app:estimates_pj}. Then, the definition of the $\mathrm{N^{th}}$-order IF estimate follows:  
\begin{definition} \label{def:FSSTN_IF_estimate}
Let $f \in L^2(\mathbb{R})$, the $N^{th}$-order local complex IF estimate $\tilde{\omega}_{\eta,f}^{[N]}$ at time $t$ and frequency $\eta$ is defined by:
\begin{equation}\tilde{\omega}_{\eta,f}^{[N]}(t,\eta) = 
\begin{cases}
\tilde{\omega}_{f} (t,\eta)  + \sum \limits_{k=2}^{N} \tilde{q}_{\eta,f}^{[k,N]} (t,\eta) \left(-x_{k,1}(t,\eta)\right),  \text{~~~if} ~V^{g}_{f}(t,\eta) \ne 0  \notag\\
\text{\hspace*{6.5cm} and}~\partial_{\eta}x_{j,j-1}(t,\eta) \ne 0~ \text{for}~j=2 \hdots N.\notag\\[0.5em]
\tilde{\omega}_f(t,\eta) \hspace*{5.3cm} \text{otherwise.}\notag
\end{cases}
\end{equation}
Then, its real part $\hat{\omega}_{\eta,f}^{[N]}(t,\eta)=\Re\{\tilde{\omega}_{\eta, f}^{[N]}(t,\eta)\}$ is the desired IF estimate. 
\end{definition}
For this estimate, we have the following approximation result:
\begin{proposition} Given a mode $f \in L^2(\mathbb{R})$ that satisfies Definition \ref{def:modeled_signal_FSSTN} with $L \le N$, then $\phi'(t) = \Re \left\{ \tilde{\omega}_{\eta,f}^{[N]}(t,\eta)\right\}$. 
\end{proposition}
\begin{IEEEproof}
From (\ref{eqn:complex_IF_nSST_general}), we have:
 \begin{align}  
 \phi'(t) &= \Re \left\{ \tilde{\omega}_{f} (t,\eta) + \sum \limits_{k=2}^{N} r_k(t) \left(-\frac{V_{f}^{t^{k-1} g}(t,\eta)}{V_{f}^{g}(t,\eta)}\right)\right\}\notag\\
&= \Re \left\{\tilde{\omega}_{f} (t,\eta) + \sum \limits_{k=2}^{N} r_k(t) \left(-x_{k,1}(t,\eta)\right)\right\}\notag\\
 &= \Re \left\{\tilde{\omega}_{f} (t,\eta) + \sum \limits_{k=2}^{N} \tilde{q}_{\eta,f}^{[k,N]} (t,\eta) \left(-x_{k,1}(t,\eta)\right)\right\}. 
\end{align}
Let us put $\tilde{\omega}_{\eta,f}^{[N]}(t,\eta) = \tilde{\omega}_{f} (t,\eta) + \sum \limits_{k=2}^{N} \tilde{q}_{\eta,f}^{[k,N]} (t,\eta) \left(-x_{k,1}(t,\eta)\right)$, we obtain $\phi'(t) = \Re \left\{ \tilde{\omega}_{\eta,f}^{[N]}(t,\eta)\right\}$, which ends the proof. 
\end{IEEEproof}

\subsection{Efficient Computation of Modulation Operators}
The local modulation operators $\tilde{q}_{\eta,f}^{[k,N]}$ defined in Proposition \ref{propo:estimates_pj} should not be computed by approximating partial derivatives 
by means of discrete differentiation, since this would generate numerical instability especially in the presence of noise. Therefore, to deal with this issue, 
we remark that these modulation operators can instead be computed analytically as functions of different STFTs. 
This is illustrated for $N = 4$ through the following proposition:   
 
\begin{proposition} \label{propo:operators_efficient} Let $f \in L^2(\mathbb{R})$, the modulation operators $\tilde{q}_{\eta,f}^{[k,N]}$ for $N = 4$ and $k=2, 3, 4$ can be expressed as:

\begin{align}
&\tilde{q}_{\eta,f}^{[4,4]} = G_{4} \left(V_{f}^{t^{0 \hdots 6}g}, V_{f}^{t^{0 \hdots 3}g'}\right),\notag\\
&\tilde{q}_{\eta,f}^{[3,4]} =  G_{3} \left(V_{f}^{t^{0 \hdots 4}g}, V_{f}^{t^{0 \hdots 2}g'}\right) - \tilde{q}_{\eta,f}^{[4,4]} G_{3,4} \left(V_{f}^{t^{0 \hdots 5}g}\right),\notag\\
& \tilde{q}_{\eta,f}^{[2,4]} = G_{2} \left(V_{f}^{t^{0 \hdots 2}g}, V_{f}^{t^{0 \hdots 1}g'}\right) - \tilde{q}_{\eta,f}^{[3,4]} G_{2,3} \left(V_{f}^{t^{0 \hdots 3}g}\right)- \tilde{q}_{\eta,f}^{[4,4]}G_{2,4} \left(V_{f}^{t^{0 \hdots 4}g}\right),\notag
\end{align}
where $G_{k} \left(V_{f}^{t^{0 \hdots m}g}, V_{f}^{t^{0 \hdots n}g'}\right)$ is a function of $V_{f}^{t^{l}g}$ for $l = 0,\hdots,m$ and $V_{f}^{t^{l}g'}$ for $l = 0,\hdots,n$ while $G_{k,j} \left(V_{f}^{t^{0 \hdots m}g}\right)$ is associated with coefficient $\tilde{q}_{\eta,f}^{[j,N]}$ in the computation of $\tilde{q}_{\eta,f}^{[k,N]}$ for $k \ne j$.        

Also, we recall that the fourth order IF estimate can be written as:    
\begin{align}
& \tilde{\omega}_{\eta,f}^{[4]}(t,\eta) = \tilde{\omega}_{f}(t,\eta) + \tilde{q}_{\eta,f}^{[2,4]} (t,\eta) \left(-x_{2,1}(t,\eta)\right) + 
\tilde{q}_{\eta,f}^{[3,4]}(t,\eta) \left(-x_{3,1}(t,\eta) \right) + \tilde{q}_{\eta,f}^{[4,4]}(t,\eta) \left(-x_{4,1}(t,\eta)\right). \notag
\end{align}
\end{proposition}

The proof of Proposition \ref{propo:operators_efficient} is available in Appendix \ref{app:operators_efficient} where explicit forms for $G_{k}$ and $G_{k,j}$ are given. 
\begin{remark} We first note that when $N=2$, i.e. by neglecting $\tilde{q}_{\eta,f}^{[3,4]}$ and $\tilde{q}_{\eta,f}^{[4,4]}$ corresponding to orders $3$ and $4$, the second-order IF estimate $\tilde{\omega}_{\eta,f}^{[2]}(t,\eta)$ defined in Proposition \ref{propo:FSST2_local_operator_frequency} is found again. Secondly, it is clear that the number of STFTs used to compute $\tilde{q}_{\eta,f}^{[4,4]}$ is 11, namely $V_{f}^{t^{l}g}$ for $l = 0,\hdots,6$, and $V_{f}^{t^{l}g'}$ for $l = 0,\hdots,3$. Finally, generalizing the procedure detailed in 
the proof of Proposition \ref{propo:operators_efficient} to any  $N$,  one obtains that $\tilde{q}_{\eta,f}^{[N,N]}$ can be computed by means of $3N-1$ STFTs, namely $V_{f}^{t^{l}g}$ for $l = 0,\hdots,2N-2$, and $V_{f}^{t^{l}g'}$ for $l = 0,\hdots,N-1$. 
\end{remark}

\subsection{Nth-order STFT-based SST (FSSTN)}
As for FSST2, the $N^{th}$-order FSST (FSSTN) is defined by replacing $\hat{\omega}_f(t,\eta)$  by $\hat{\omega}_{\eta,f}^{[N]}(t,\eta)$ in (\ref{eqn:SST_operator}):
\begin{definition} \label{def:FSSTN}
Given $f \in L^2({\mathbb{R})}$ and a real number $\gamma > 0 $, one defines the FSSTN operator with threshold $\gamma$ as:
 \begin{align}  
  &T_{N, f}^{g,\gamma}(t,\omega) = \dfrac{1}{g^*(0)} \int_{\{\eta,|V_f^g(t,\eta)|>\gamma\}} V_f^g(t,\eta) \delta \left(\omega-\hat{\omega}_{\eta,f}^{[N]}(t,\eta) \right)d\eta. \notag
 \end{align}
\end{definition}
Finally, the modes of the MCS can be reconstructed by replacing $T_{f}^{g,\gamma}(t,\omega)$ by $T_{N, f}^{g,\gamma}(t,\omega)$ in (\ref{eqn:reconstruction_formula_SST}). 

\section{Numerical Analysis of the Behavior of STFT-based SST}
\label{sec:numerical_implementation}
This section presents numerical investigations to illustrate the improvements brought by our new technique in comparison with the standard reassignment method (RM) or existing STFT-based SSTs (FSST and FSST2) on both simulated and real signals. 
For that purpose, let us first consider a simulated MCS composed of two AM-FM components:
\begin{align*}
 f(t) &= f_1(t) + f_2(t) = A_1(t) e^{i2\pi\phi_1(t)} + A_2(t) e^{i2\pi\phi_2(t)},  \notag
 \end{align*}
 with $A_k(t)$ and $\phi_k(t)$  defined on $[0,1]$, for $k=1,2$, by:
\begin{align*}
 A_1(t) &= \exp \left(2(1-t)^3 + t^4\right),~A_2(t) = 1+5t^2 + 7(1-t)^6 ~\text{and}~\notag\\
 \phi_1(t) &= 50t+30t^3-20(1-t)^4,~ \phi_2(t) = 340t-2\exp\left(-2(t-0.2)\right)\sin\left(14\pi(t-0.2)\right). \notag
\end{align*}

\begin{figure}[!htb]
\begin{minipage}[b]{\linewidth}
\centering
\centerline{\includegraphics[width=0.4\linewidth]{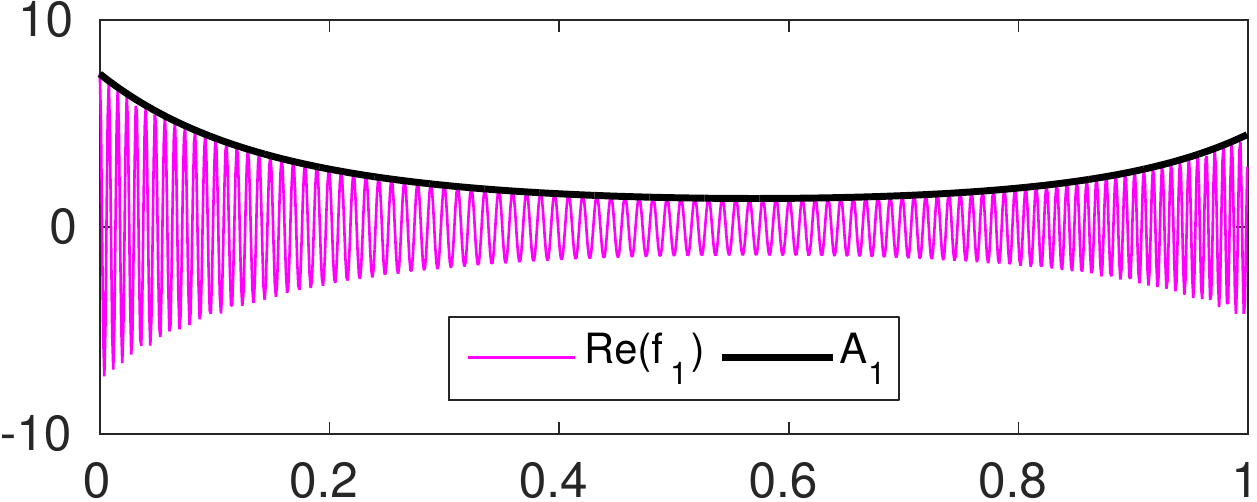}}
\centerline{(a)} 
\end{minipage}
\begin{minipage}[b]{\linewidth}
\centering
\centerline{\includegraphics[width=0.4\linewidth]{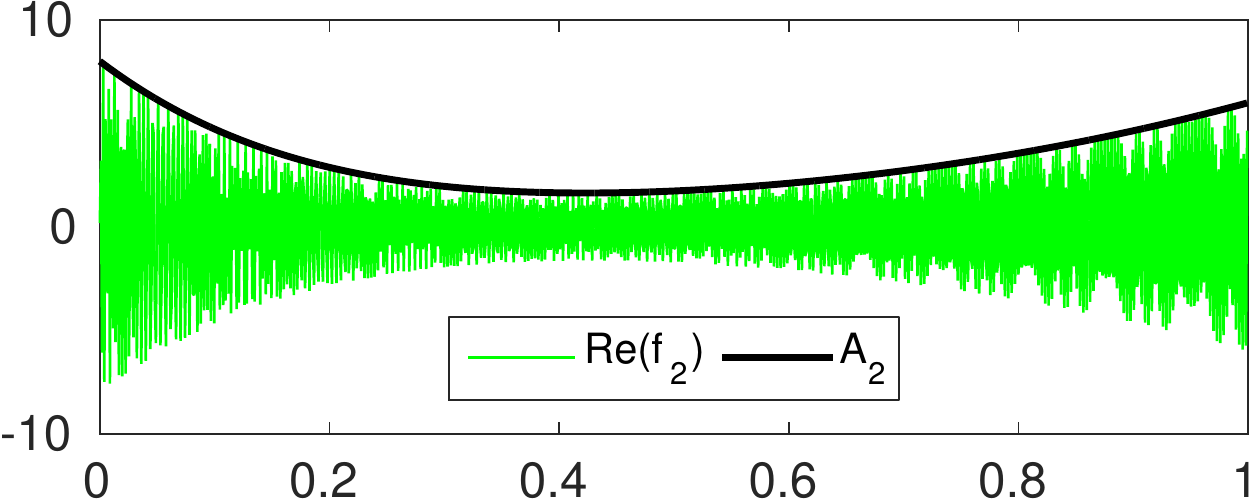}}
\centerline{(b)} 
\end{minipage}
\begin{minipage}[b]{\linewidth}
\centering
\centerline{\includegraphics[width=0.4\linewidth]{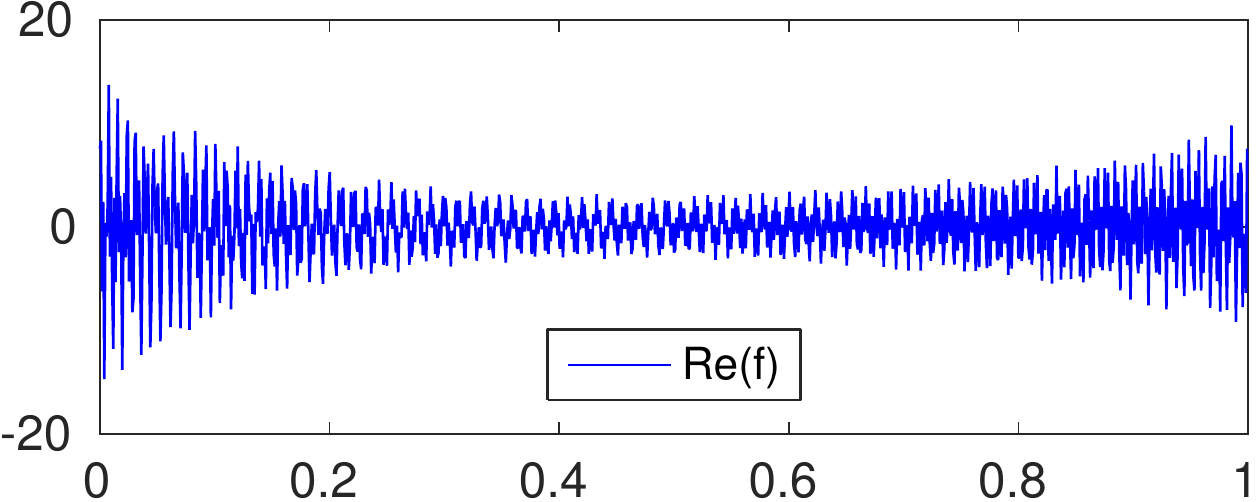}}
\centerline{(c)} 
\end{minipage}
\caption{(a) and (b): real part of $f_1$ and $f_2$ respectively with $A_1$ and $A_2$ superimposed; (c): real part of $f$.}
\label{fig:Fig1}
\end{figure}

Note that $f_1$ is a polynomial chirp that satisfies Definition \ref{def:modeled_signal_FSSTN} with $L=N=4$, while $f_2$ is a damped-sine function containing very strong nonlinear sinusoidal frequency modulations and high-order polynomial amplitude modulations. In our simulations, $f$ is sampled at a rate $M = 1024$ Hz on $[0,1]$. In Figures \ref{fig:Fig1}(a) and (b), we display the real part of $f_1$ and $f_2$ along with their amplitudes, and, in Figure \ref{fig:Fig1} (c), the real part of $f$. 

\begin{figure}[!htb]
\begin{minipage}[b]{\linewidth}
\centering
\centerline{\includegraphics[width=0.45\linewidth]{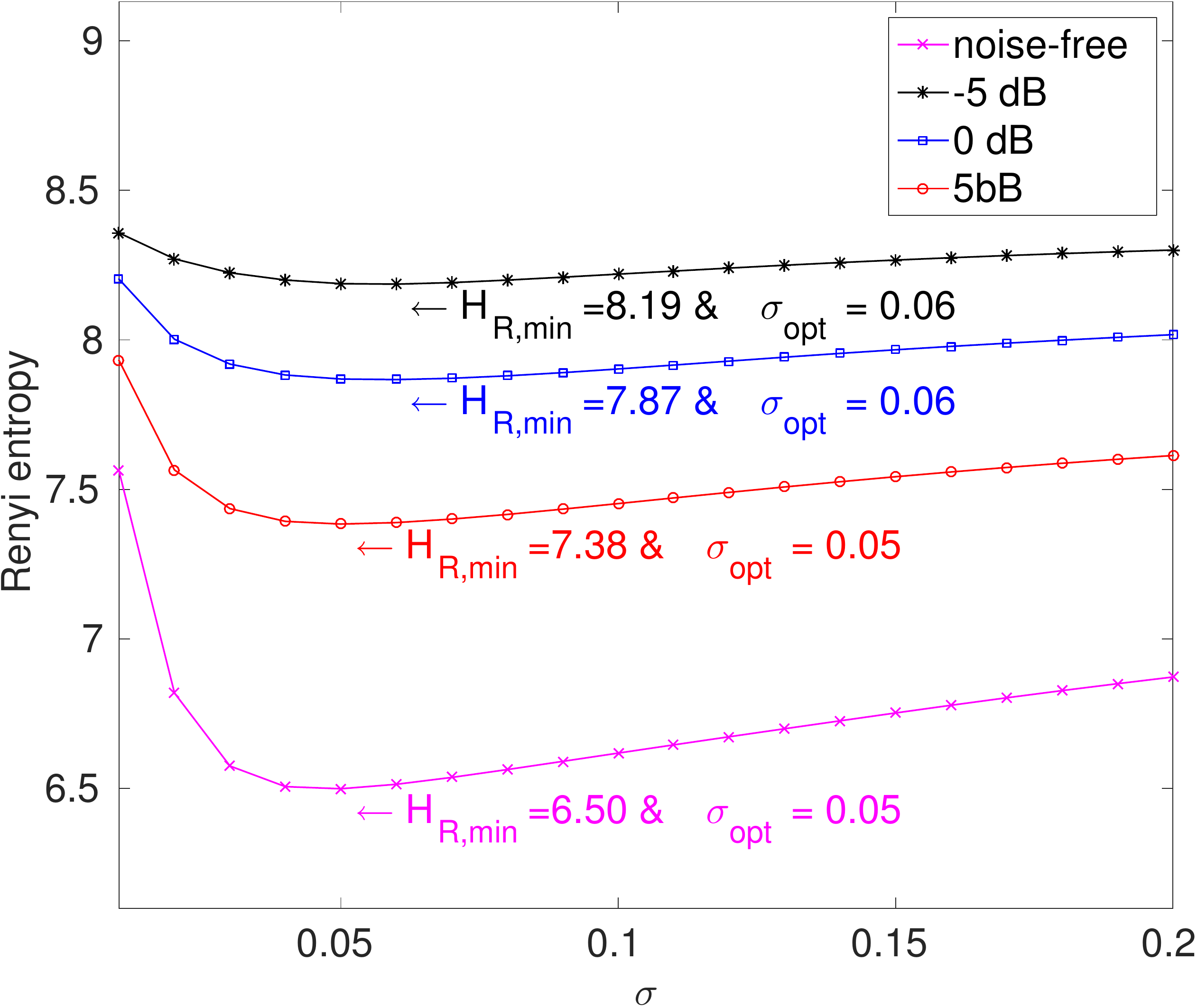}}
\end{minipage}
\caption{Evolution of R\'enyi entropies ($H_R$) with respect to $\sigma$ either in the noise-free, 5 dB, 0 dB or -5 dB cases.}
\label{fig:Fig2}
\end{figure}

The STFT of $f$ is then computed with the $L^1-$normalized Gaussian window 
$g(t) = \sigma^{-1}e^{-\pi \frac{t^2}{\sigma^2}}$, where $\sigma$ is optimal in some sense as explained hereafter. 
One of the well-known issues regarding the use of STFT to analyze signals is the choice of an appropriate Gaussian window length to allow for a good trade-off between time and 
frequency localization. In the synchrosqueezing context, the choice of analysis window for the STFT has a strong impact on the accuracy of mode reconstruction: to use an inappropriate window may lead to the failure of ridge extraction and then of mode retrieval. To deal with this issue, a widely used approach is to measure the concentration of 
STFT which then allows us to pick the `optimal' window length as the one associated with the most concentrated representation. For that purpose, a relevant work is \cite{Stankovic2001}, in which the concentration of the STFT is measured by means of R\'enyi entropy:
\begin{align} \label{eqn:Renyi}
H_R(\sigma) = \frac{1}{1-\alpha} \log_2 \left ( \frac{\int \int_{\mathbb{R}^2} |V_f^g(t,\eta)|^\alpha d\eta dt}{\int \int_{\mathbb{R}^2} |V_f^g(t,\eta)|d\eta dt} \right ), 
\end{align}
with integer orders $\alpha > 2$ being recommended. The larger the R\'enyi entropy, the less concentrated the STFT. The optimal window length parameter is thus determined as: $ \sigma_{opt} = \text{arg}\min \limits_{\sigma} \left( H_R(\sigma) \right)$. In Figure \ref{fig:Fig2}, we display the evolution of R\'enyi entropy $(\alpha = 3)$ with respect to $\sigma$ for the signal $f$ introduced above at different noise levels (noise-free, -5, 0, 5 dB), which leads to an optimal value in each case, relatively stable with the noise level.   

\begin{figure*}[!htb]
\begin{minipage}[b]{0.33\linewidth}
\centering
\includegraphics[width=.818\linewidth,right]{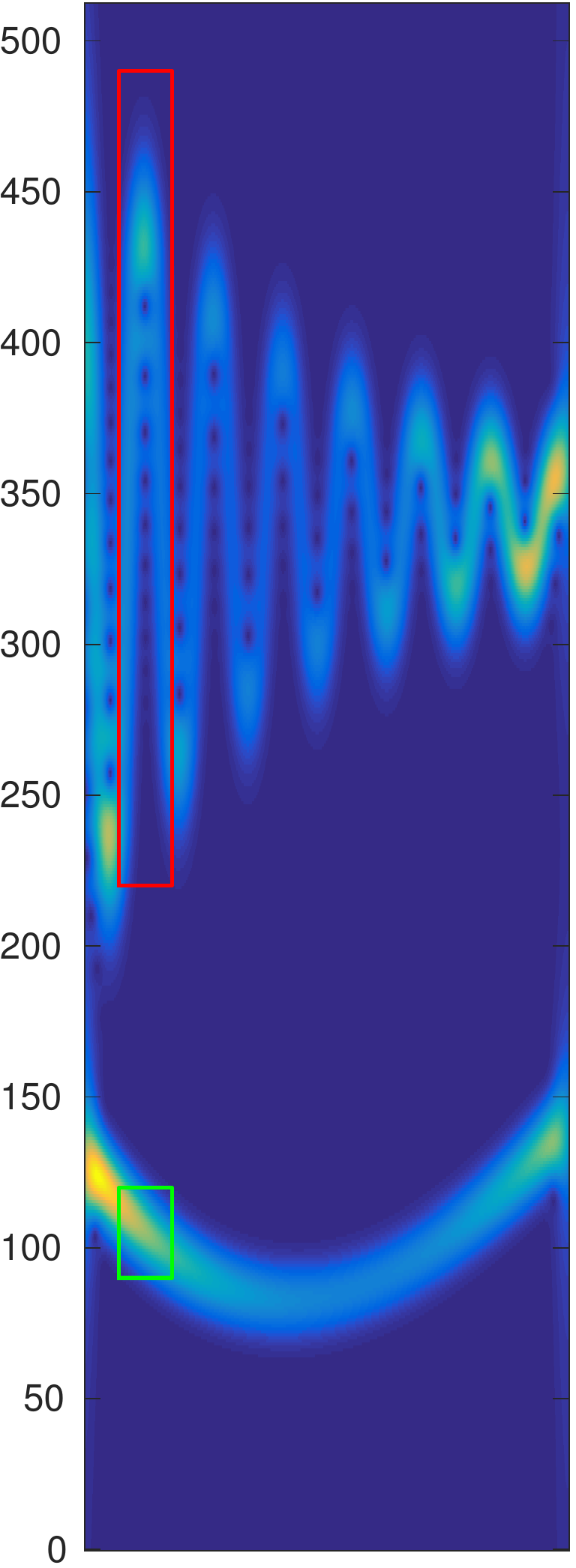}
\centerline{~~~~~~~~~~~(a) STFT} 
\end{minipage}
\begin{minipage}[b]{0.33\linewidth}
\centering
\centerline{\includegraphics[width=.6\linewidth]{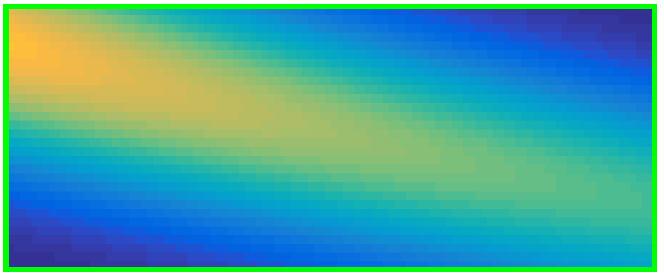}}
\centerline{(b) STFT} 
\centerline{\includegraphics[width=.6\linewidth]{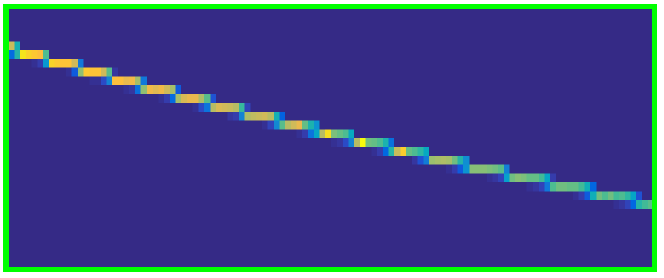}}
\centerline{(c) RM} 
\centerline{\includegraphics[width=.6\linewidth]{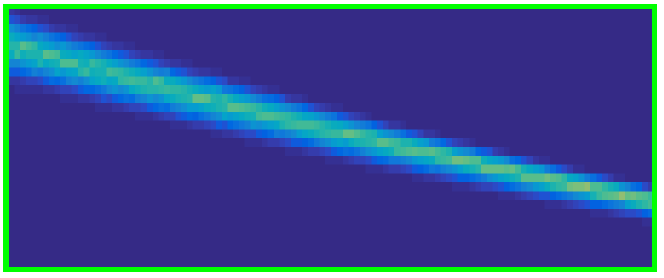}}
\centerline{(d) FSST} 
\centerline{\includegraphics[width=.6\linewidth]{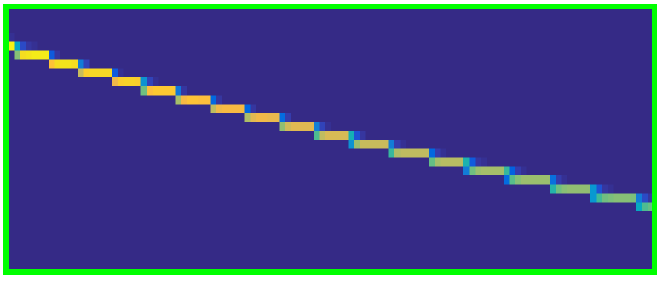}}
\centerline{(e) FSST2} 
\centerline{\includegraphics[width=.6\linewidth]{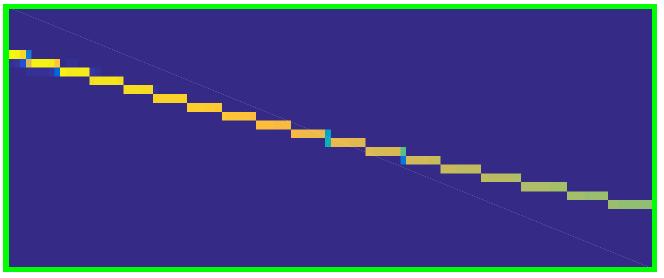}}
\centerline{(f) FSST3} 
\centerline{\includegraphics[width=.6\linewidth]{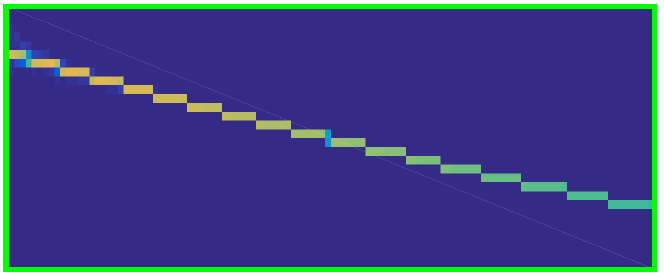}}
\centerline{(g) FSST4} 
\end{minipage}
\begin{minipage}[b]{0.33\linewidth}
\centering
\includegraphics[width=.6\linewidth,left]{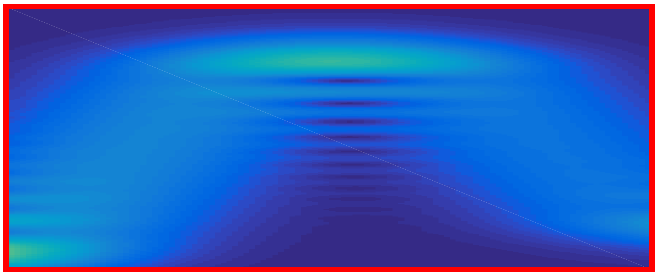}
\centerline{(h) STFT~~~~~~~~~~~~~~~~} 
\includegraphics[width=.6\linewidth,left]{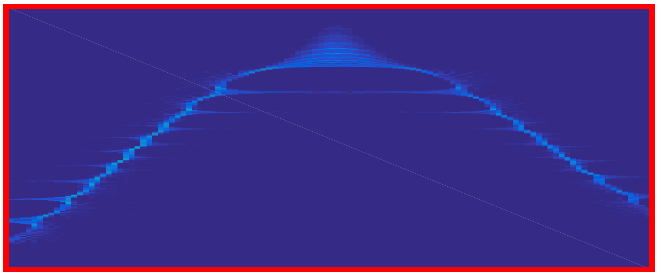}
\centerline{(i) RM~~~~~~~~~~~~~~~~} 
\includegraphics[width=.6\linewidth,left]{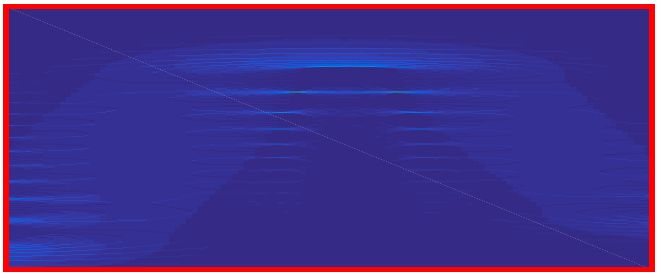}
\centerline{(j) FSST~~~~~~~~~~~~~~~~} 
\includegraphics[width=.6\linewidth,left]{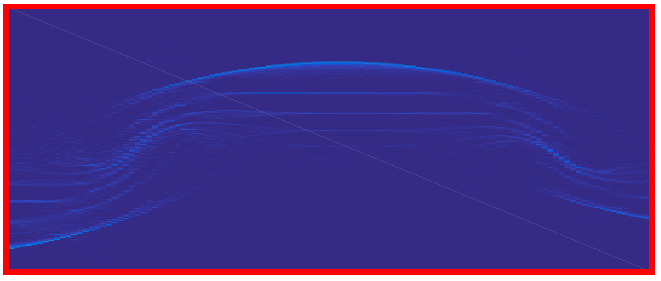}
\centerline{(k) FSST2~~~~~~~~~~~~~~~~} 
\includegraphics[width=.6\linewidth,left]{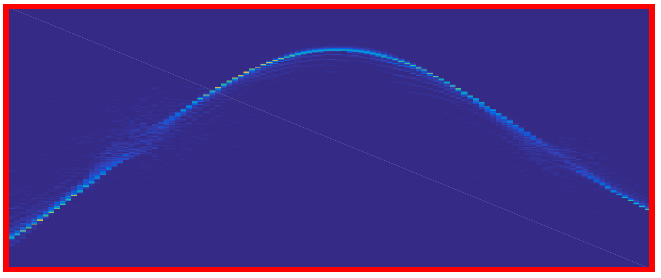}
\centerline{(l) FSST3~~~~~~~~~~~~~~~~} 
\includegraphics[width=.6\linewidth,left]{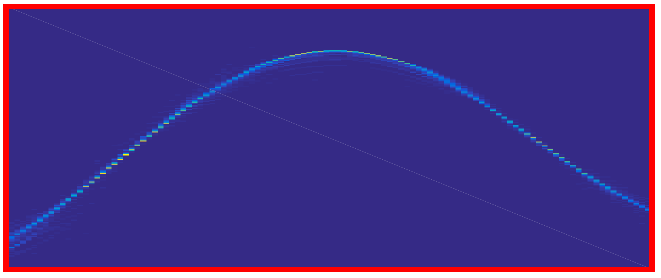}
\centerline{(m) FSST4~~~~~~~~~~~~~~~~} 
\end{minipage}
\caption{\textit{Right column panel}, (a): modulus of the STFT of  $f$. \textit{Middle column panel}, (b): STFT of a small TF patch corresponding to mode $f_1$ (delimited by green segments)  extracted from (a); (c) RM carried out on the STFT shown in (b); from (d) to (g),  same as (c) but using respectively FSST, FSST2, FSST3, FSST4. \textit{Left column panel},  same as \textit{middle column panel} but for $f_2$.}
\label{fig:Fig3}
\end{figure*}

Having determined the optimal $\sigma$, we display,  in the noise-free context, the STFT of $f$ on the left of Figure  \ref{fig:Fig3}. Then, on the right of this figure, close-ups of the STFT itself are depicted, along with  reassigned versions of STFT either given by the reassignment method (RM) or FSST and variants,  
all mentioned in this paper. For the sake of consistency, we recall that RM corresponds to the reassignment of the spectrogram through \cite{Auger1995}:
\begin{align} 
\label{eqn:RM_operator}
 &RM_{f}^{g}(t,\omega) = \int \int_{\mathbb{R}^2} |V_f^g(\tau,\eta)|^2 \delta \left(\omega-\hat{\omega}_f(\tau,\eta) \right) \delta \left(t-\hat{\tau}_f(\tau,\eta)\right) d \eta d \tau. \notag
\end{align}
It behaves well with frequency modulation, but does not allow for mode reconstruction. 

Analyzing these close-ups, we remark that, as expected, FSST2 leads to a relatively sharp TF representation for $f_1$, very similar to the one given by  RM and much better 
than that corresponding to FSST.  
However, all these methods fail to reassign the STFT of $f_2$ correctly, especially where the IF of that mode has a non negligible curvature $\phi'''_2(t)$.  
In contrast, the TF reassignment of the STFT of $f_2$ provided by FSST3 or FSST4 is much sharper at these locations.
Looking at what happens for mode $f_1$  also tells us that, FSST3 and FSST4 seems to behave very similarly to FSST2 or RM in terms of the 
sharpness of the representation. However, as we shall see later, the accuracy of the representation is improved by using one of the former two methods. 
Finally, note that since $f_1$ obeys Definition  \ref{def:modeled_signal_FSSTN}, the IF estimate  used in FSST4 is exact for that mode which results 
in perfect reassignment of the STFT.

For a better understanding of the performance improvements brought by the use of FSST3 and FSST4 over other studied methods, the following section first 
introduces a quantitative comparison of all these techniques from the angle of energy concentration of TF representations, and then a measure of their accuracy 
by means of the Earth mover's distance (EMD).        

\subsection{Evaluation of TF Concentration}

\begin{figure}[!htb]
\begin{minipage}[b]{0.48\linewidth}
\centering
\centerline{\includegraphics[width=0.75\linewidth]{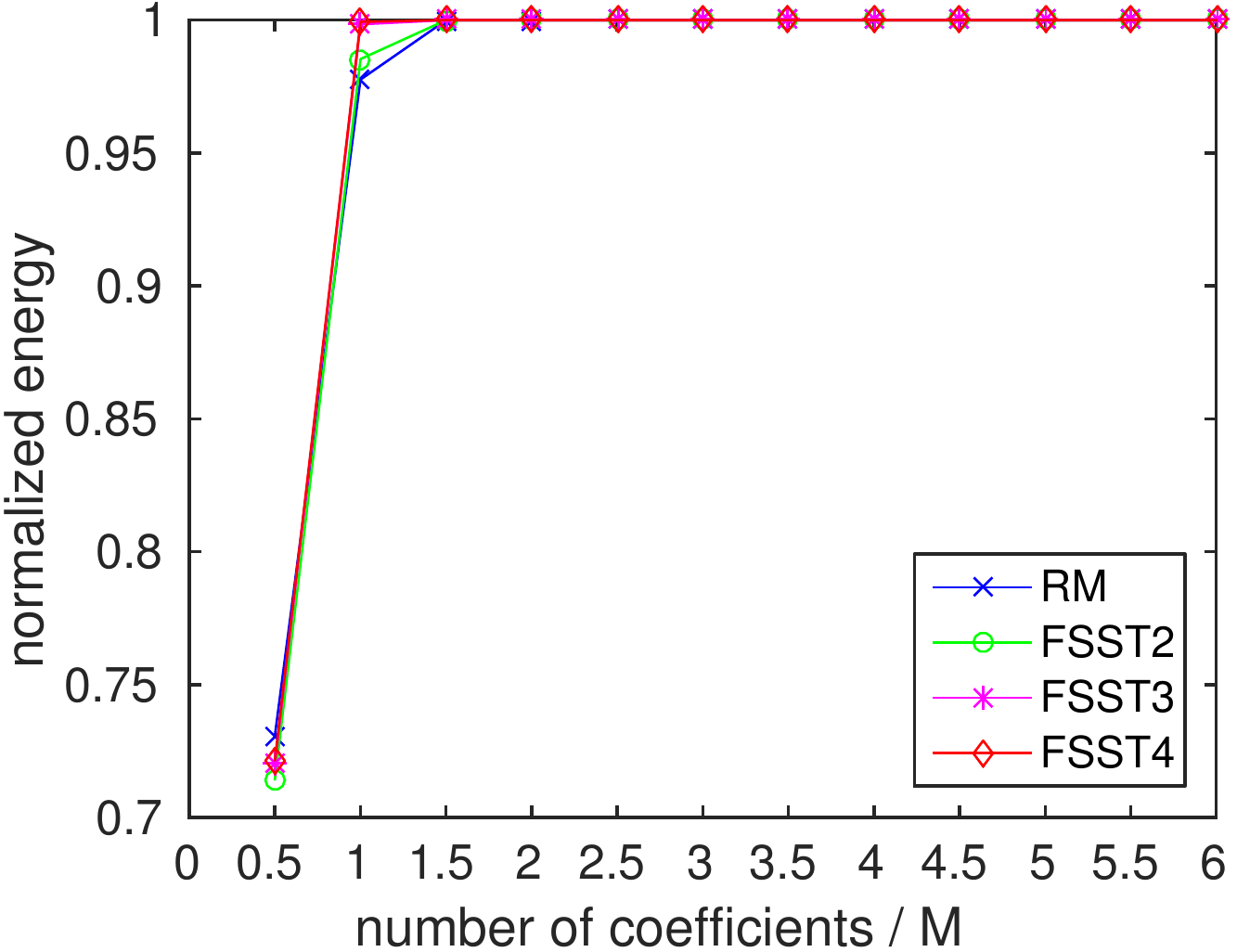}}
\centerline{(a)} 
\end{minipage}
\begin{minipage}[b]{0.48\linewidth}
\centering
\centerline{\includegraphics[width=0.75\linewidth]{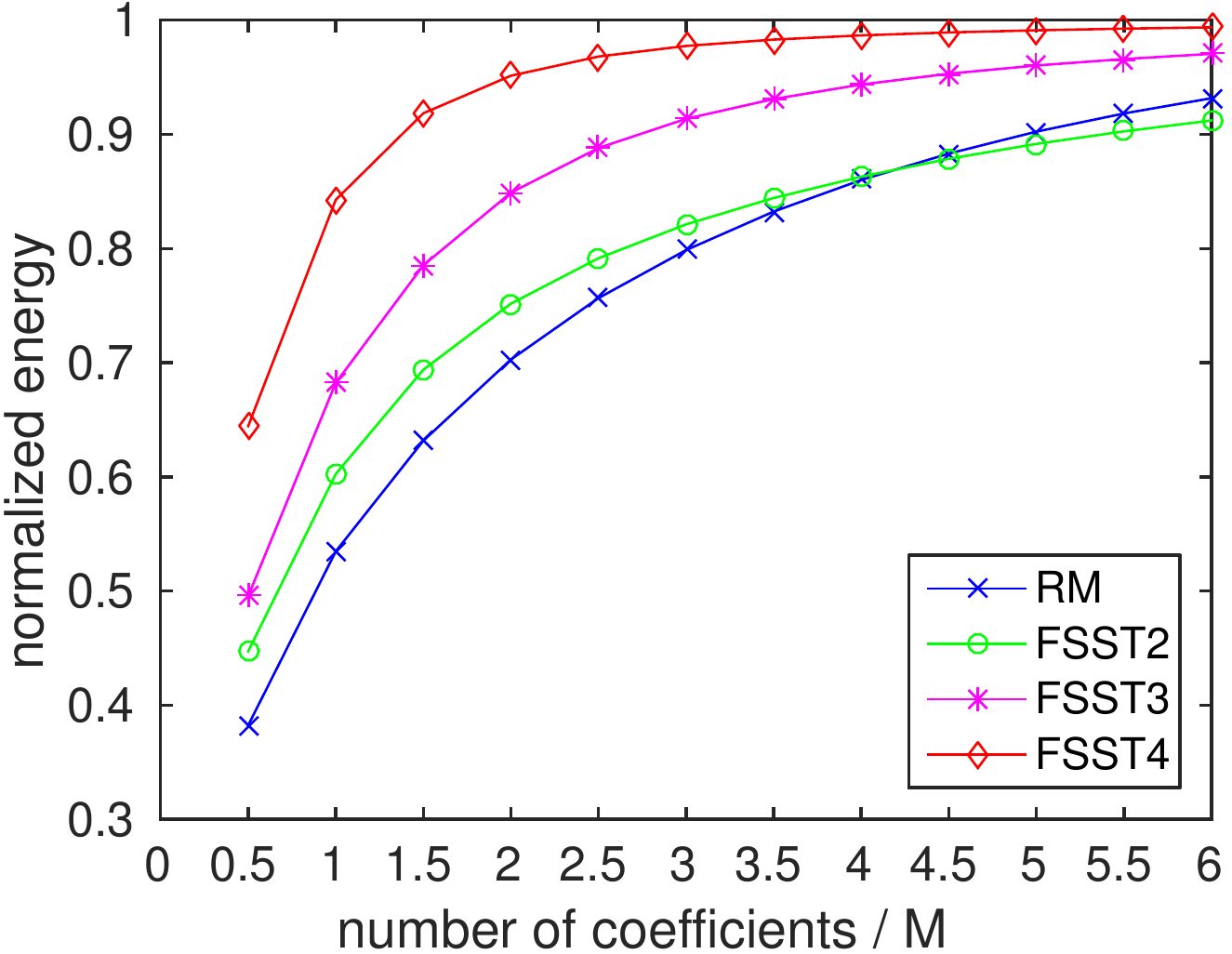}}
\centerline{(b)} 
\end{minipage}
\caption{(a) Normalized energy as a function of the number of sorted associated coefficients for  $f_1$; 
(b): same as (a) but for $f_2$.}
\label{fig:Fig4}
\end{figure}

To evaluate the performance of the different techniques regarding TF concentration, we first use a method introduced in \cite{Oberlin2015}. The goal of this method is to measure the energy concentration by considering the proportion of the latter contained in the first nonzero coefficients associated with the highest amplitudes, which we call {\it normalized energy} in the sequel: the faster it increases towards 1 with the number of coefficients involved, the more concentrated the TF representation. In Figure \ref{fig:Fig4} (a), we depict the normalized energy corresponding to the reassignment of the STFT of  $f_1$ using different techniques, 
with respect to the number of coefficients kept divided by the length of $f_1$ (which corresponds to the sampling rate $M$ in our case). 
Since we consider only one mode, a good representation has to have its  energy mostly contained in the first $M$ coefficients, which correspond to abscissa 
1 in the graph of Figure \ref{fig:Fig4} (a). From this study and from this signal, it is hard to figure out the benefits of using FFST3 or FSST4 rather than the other two methods. The only thing one can check is that the energy is perfectly localized with FSST4 because $f_1$ obeys Definition \ref{def:modeled_signal_FSSTN}.
The results of the same computation carried out for mode $f_2$ are displayed in Figure  \ref{fig:Fig4} (b), showing that  the 
normalized energy is much more concentrated using FSST4 than the other methods, and that FSST3 also outperforms FSST2 and RM.    

\begin{figure}[!htb]
\begin{minipage}[b]{0.48\linewidth}
\centering
\centerline{\includegraphics[width=0.75\linewidth]{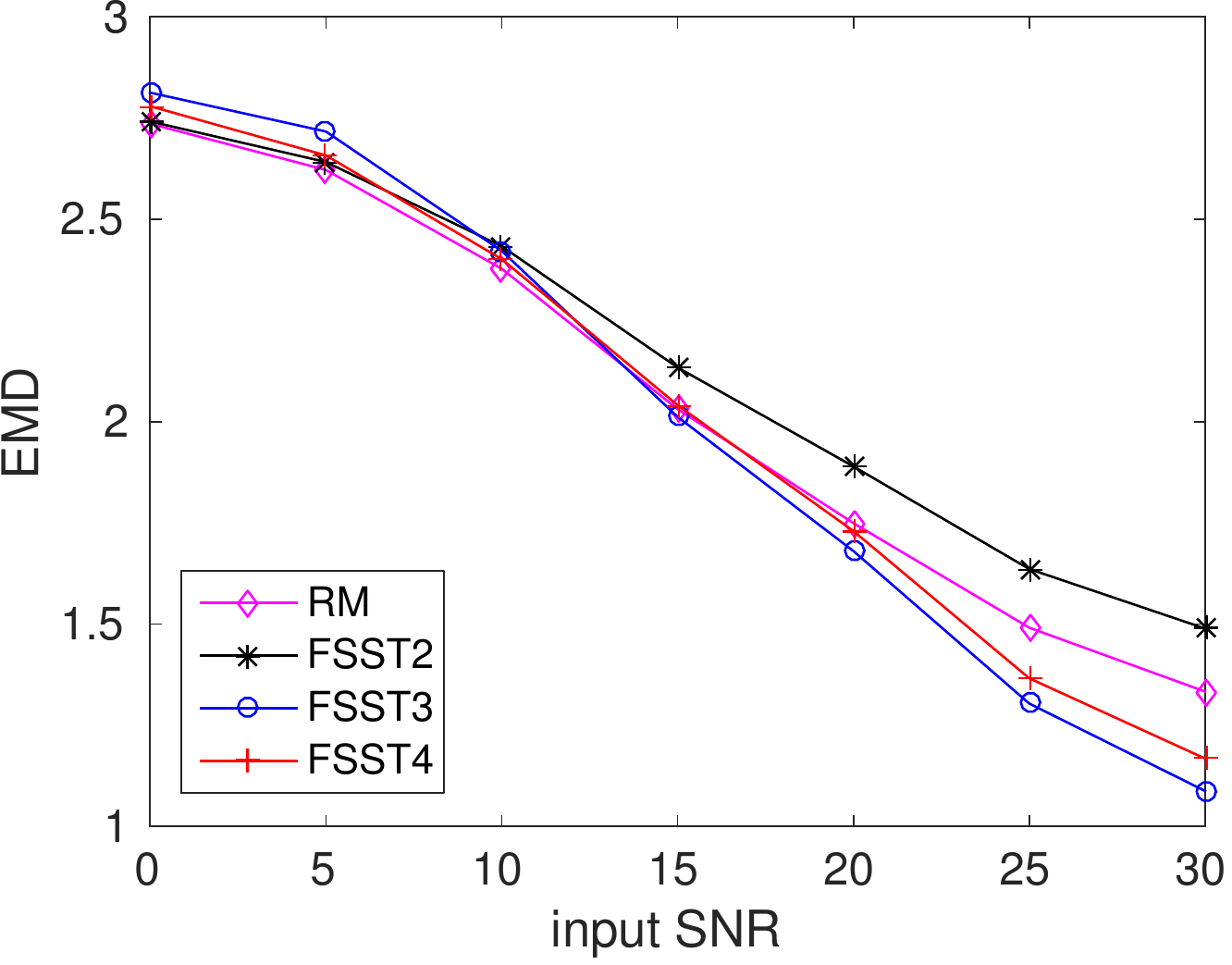}}
\centerline{(a)} 
\end{minipage}
\begin{minipage}[b]{0.48\linewidth}
\centering
\centerline{\includegraphics[width=0.75\linewidth]{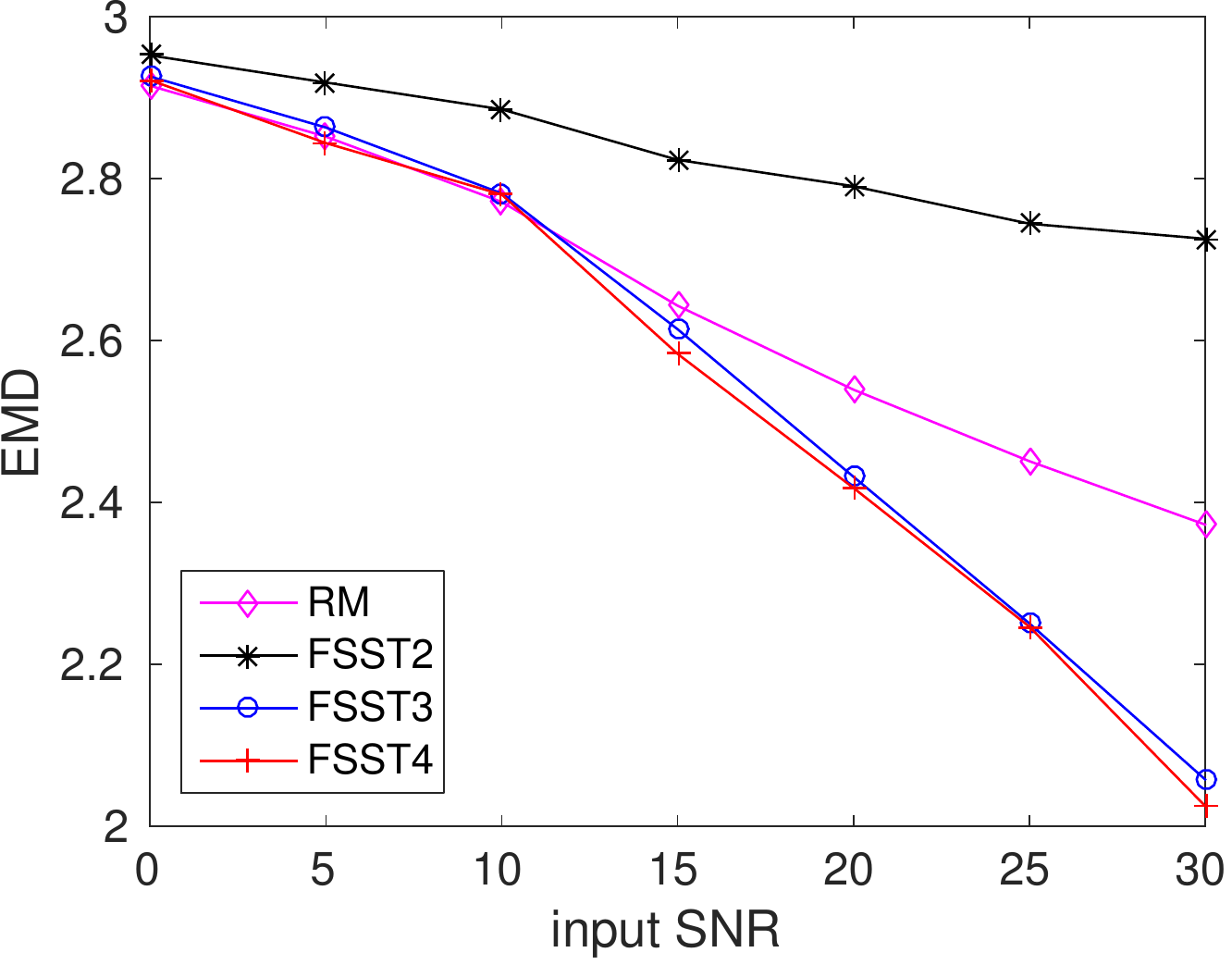}}
\centerline{(b)} 
\end{minipage}
\caption{(a): EMD corresponding to different TF representations of $f_1$ either given by RM, FSST2, FSST3 or FSST4; (b): same as (a) but for $f_2$.}
\label{fig:Fig5}
\end{figure}

To study the performance of the TF representations in the presence of noise, we consider a noisy signal, 
where  the noise level is measured by the Signal-to-Noise Ratio (SNR): 
\begin{align} \label{eqn:SNR_definition}
\text{SNR}_{\text{intput}}[\text{dB}]  = 20\log_{10} \frac{\text{std}(f)}{\text{std} (\zeta)}, ~\text{where~std is the standard deviation, }
\end{align}
and $\zeta(t)$ is the white Gaussian noise added. 
To compute the normalized energy as illustrated in Figure \ref{fig:Fig4}, though quite informative, does not deliver any insight into 
the accuracy of the reassigned transforms. The latter can alternatively be quantified  by measuring the dissimilarity between the resultant TF representations 
and the ideal one by means of the Earth mover's distance (EMD), a procedure already used in the synchrosqueezing context in \cite{Daubechies2016}. 
More precisely, this technique consists in computing the 1D EMD between the resultant TF representations and the ideal one, for each individual time $t$, and then take the average over all $t$ to define  
the global EMD. 
A smaller EMD means a better TF representation concentration to the ground truth and less noise fluctuations. In Figures \ref{fig:Fig5} (a) and (b), 
we display, respectively for $f_1$ and $f_2$, the evolution of EMD  with respect to the noise level, for TF representations given 
either by FSST2, FSST3, FSST4 or RM. This study tells us that, at low noise level and for mode $f_1$, FSST3 and FSST4 
are more accurate than the other studied methods. Note that this is something that could not be derived by the previous study on the normalized energy. 
The same investigations but for mode $f_2$ confirms the interest of using FSST3 or FSST4 to reassign the STFT of a mode with IF exhibiting strong curvature.
 Note that the benefits of using the proposed new methods remain important even at high noise level.

\subsection{Evaluation of Mode Reconstruction Performance}

As discussed above, the  variants of FSST proposed in this paper leading to significantly better TF representations,  this should translate into better performance in terms of mode reconstruction. Let us first briefly recall the procedure to retrieve $f_k$  from the TF representation of $f$ given by the FSST of order $N$:        
\begin{equation} \label{eqn:reconstruction_formula_FSSTN}
 f_k(t) \approx \int_{\{\omega,|\omega-\varphi_k(t)|<d\}}T_{N, f}^{g,\gamma}(t,\omega)d\omega.
\end{equation}
Note that $\varphi_k(t)$ is the estimate of $\phi_k'(t)$ given by the ridge detector (computed by minimizing energy  (\ref{eqn:energy_functional}) in which 
$T_{f}^{g,\gamma}$ is replaced by $T_{N, f}^{g,\gamma}$),  and $d$ is an integer parameter (because the frequency resolution is here associated 
with integer location) used to compensate for the inaccuracy of this estimation and also for the errors caused by approximating the IF  by $\hat{\omega}_{\eta,f}^{[N]}(t,\eta)$. 
We first analyze the performance of the reconstruction procedure by considering the information on the ridge only, i.e. we take $d=0$. For that purpose, we measure 
the output SNR, defined by $\text{SNR}_{\text{output}} = 20\log_{10} \dfrac{\left\Vert f \right\Vert_2}{\left\Vert f_r -f  \right\Vert_2} $, where $f_r$ is the reconstructed signal, and 
$\left\Vert .  \right\Vert_2$ the $l_2$ norm. In Table \ref{tab:mode_retrieval_accuracy}, we display this output SNR for modes $f_1$, $f_2$ and also for $f$, 
using either FSST2, FSST3 or FSST4 for mode reconstruction. The improvement brought by using FSST3 and FSST4 is clear  and coherent with the previous study of 
the accuracy of the proposed new TF representations. 
\begin{table}[!htb]
\renewcommand{\arraystretch}{1.3}
\caption{Performance of mode reconstruction in the noise-free case}
\label{tab:mode_retrieval_accuracy}
\centering
\begin{tabular}{|c|c|c|c|}
\hline
\bfseries  & \bfseries FSST2 & \bfseries FSST3 & \bfseries FSST4\\
\hline
Mode $f_1$  &  17.8 & 25.7 & \bfseries 28.8\\
\hline
Mode $f_2$  & 1.73 & 3.62 & \bfseries 6.87 \\ 
\hline
MCS  $f$    & 3.57 & 5.57 & \bfseries 8.82 \\
\hline
\end{tabular}
\end{table} 

\begin{figure}[!htb]
\begin{minipage}[b]{\linewidth}
\centering
\centerline{\includegraphics[width=0.35\linewidth]{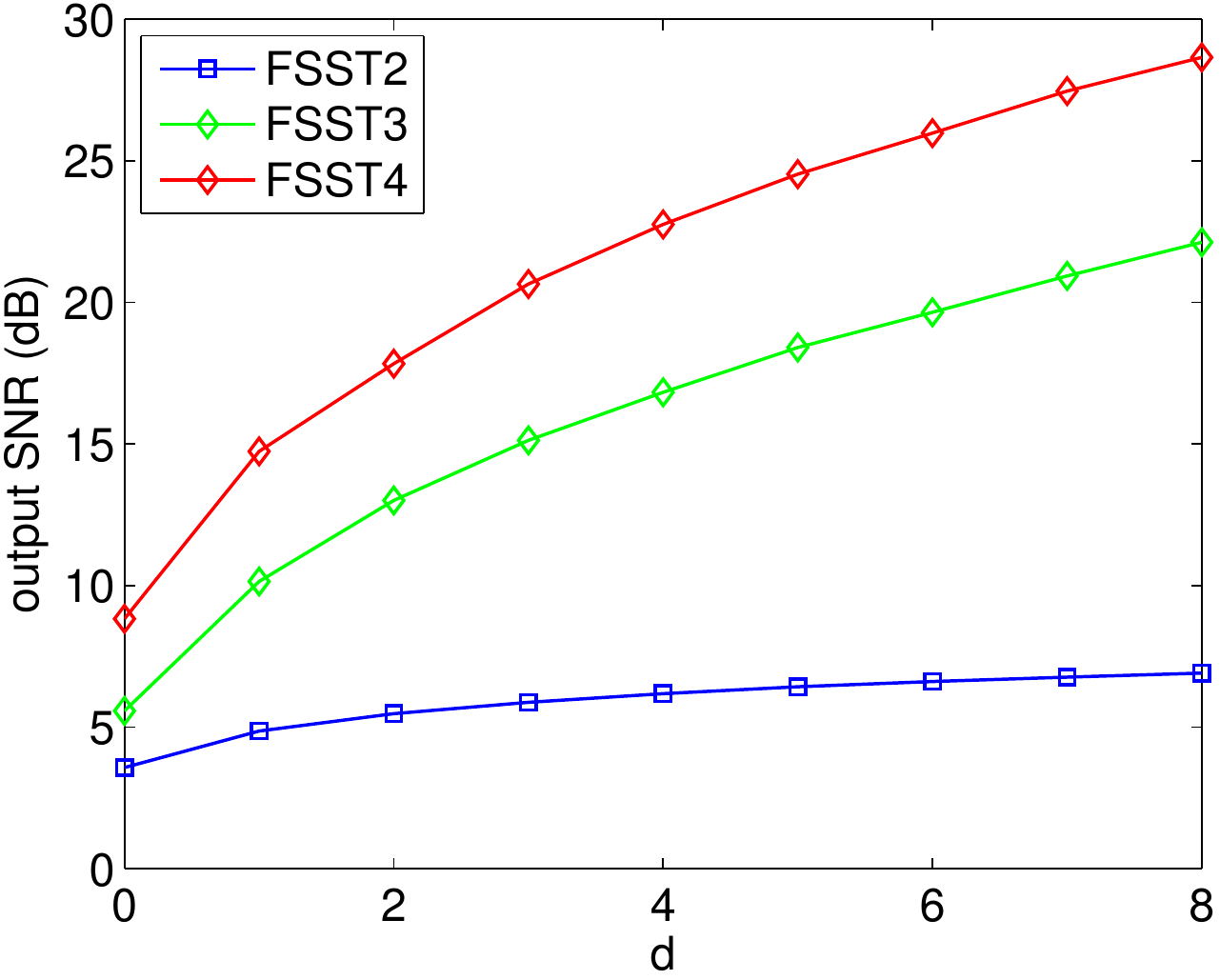}}
\end{minipage}
\caption{Reconstruction accuracy measured in SNR with respect to $d$ of the noise-free signal.}
\label{fig:Fig6}
\end{figure}

Parameter $d$  also measures how well the TF representation is concentrated around the detected ridges: if the former is well concentrated, even if one uses a small 
$d$, the reconstruction results should be satisfactory.  To measure this, we display in Figure \ref{fig:Fig6}, the output SNR corresponding to the reconstruction of  
$f$ when $d$ varies, and when the TF representation used for mode reconstruction is either FSST2, FSST3 and FSST4. From this Figure and for all tested methods, it is clear that a larger $d$ means a more accurate reconstruction of the signal. Nevertheless,  the accuracy the reconstruction using FSST2 seems to stagnate when some critical value for $d$ is 
reached, which is not the case with the other two methods: the parameter $d$ can only partly compensate for the inaccuracy of IF estimation.
For that very reason, it is crucial to use the most accurate estimate as possible which again pleads in favor of FSST3 and FSST4.

\subsection{Application to Gravitational-wave Signal}

In this section, we investigate the applicability of our new techniques for the analysis of a transient gravitational-wave signal, which was generated by the coalescence of two stellar-mass black holes. This event, called \textbf{GW150914}, was recently detected by the LIGO detector Hanford, Washington. Such a signal closely matches with waveform Albert Einstein predicted almost 100 years ago in his general relativity theory for the inspiral, the merger of a pair of black holes and the ringdown
of the resulting single black hole \cite{Abbott2016}.  
The observed signal has a length of 3441 samples in 0.21 seconds, which we pad with zeros to get a signal with $2^{12}$ samples, and  the  Gaussian window 
used in our simulations  corresponds to $\sigma = 0.05$.

\begin{figure*}[!htb]
\begin{minipage}[b]{0.24\linewidth}
\centering
\centerline{
\includegraphics[width = 0.95\linewidth, height = 4.5cm]{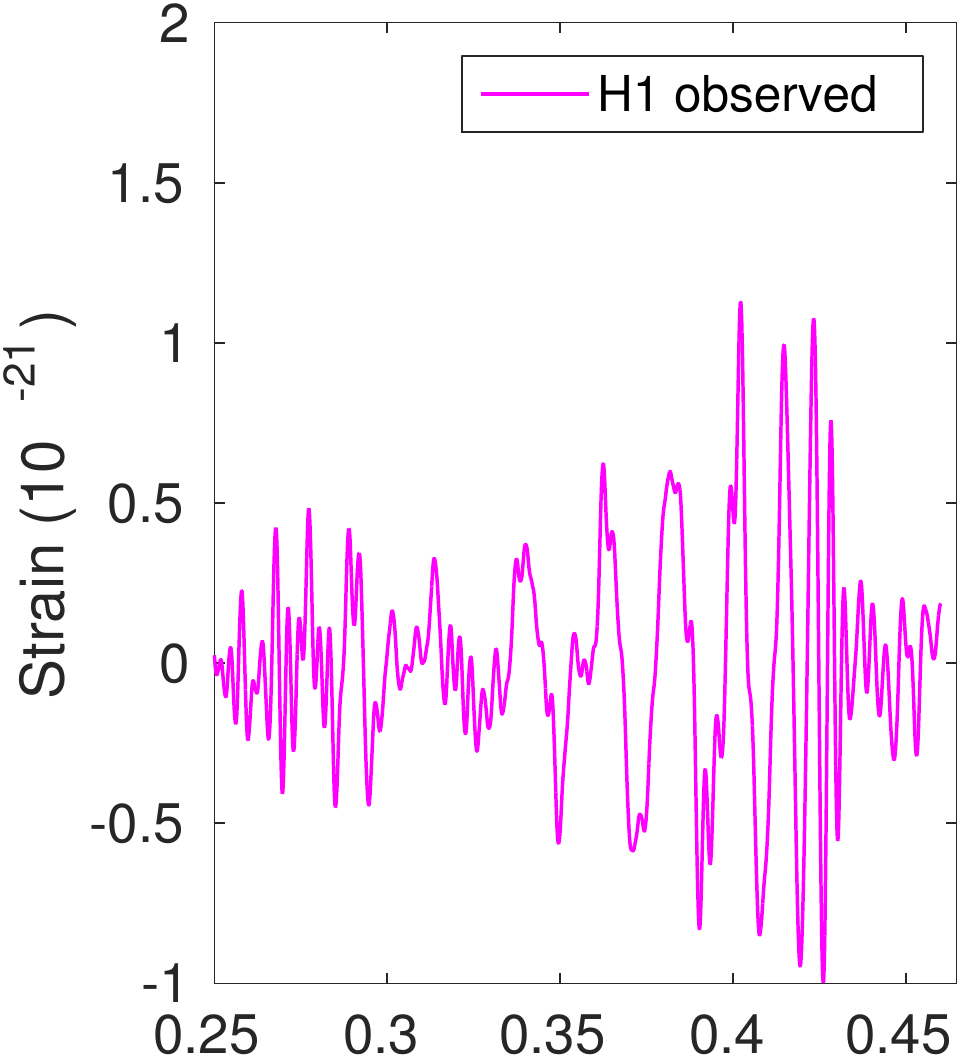}}
\centerline{(a)} 
\end{minipage}
\begin{minipage}[b]{0.24\linewidth}
\centering
\centerline{
\includegraphics[width = 0.95\linewidth, height = 4.5cm]{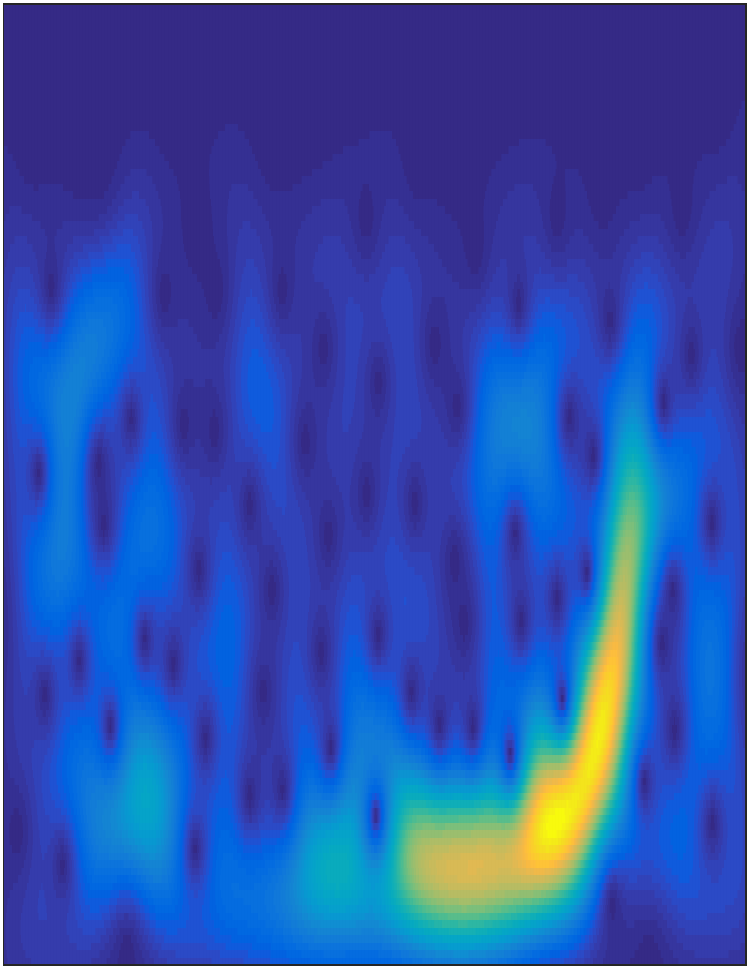}}
\centerline{(b)} 
\end{minipage}
\begin{minipage}[b]{0.24\linewidth}
\centering
\centerline{
\includegraphics[width = 0.95\linewidth, height = 4.5cm]{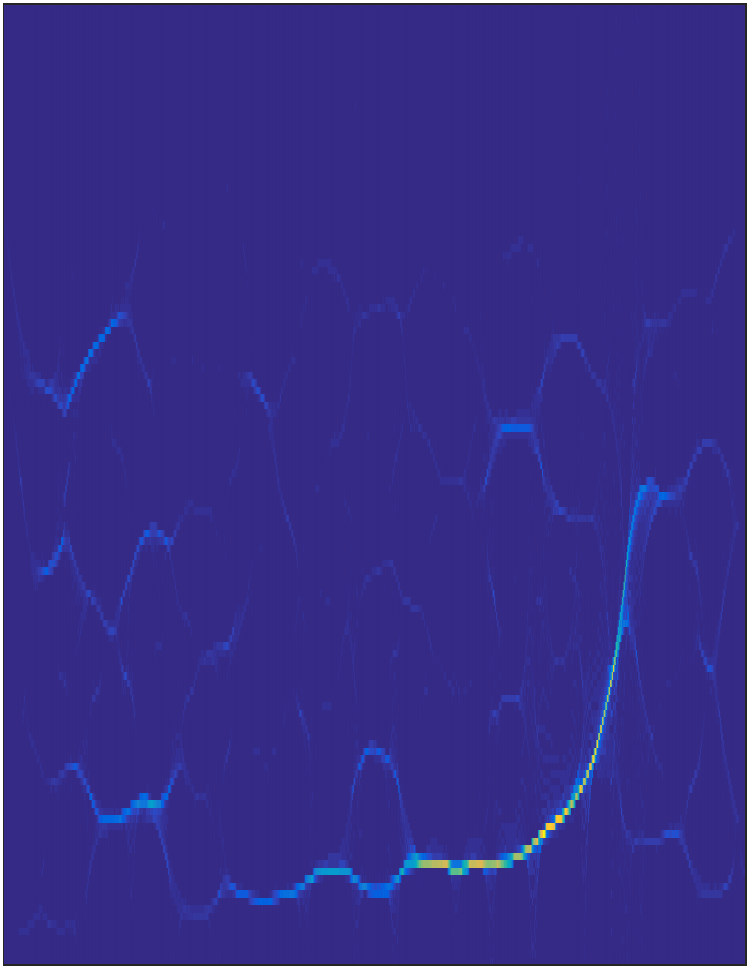}}
\centerline{(c)} 
\end{minipage}
\begin{minipage}[b]{0.24\linewidth}
\centering
\centerline{
\includegraphics[width = 0.95\linewidth, height = 4.5cm]{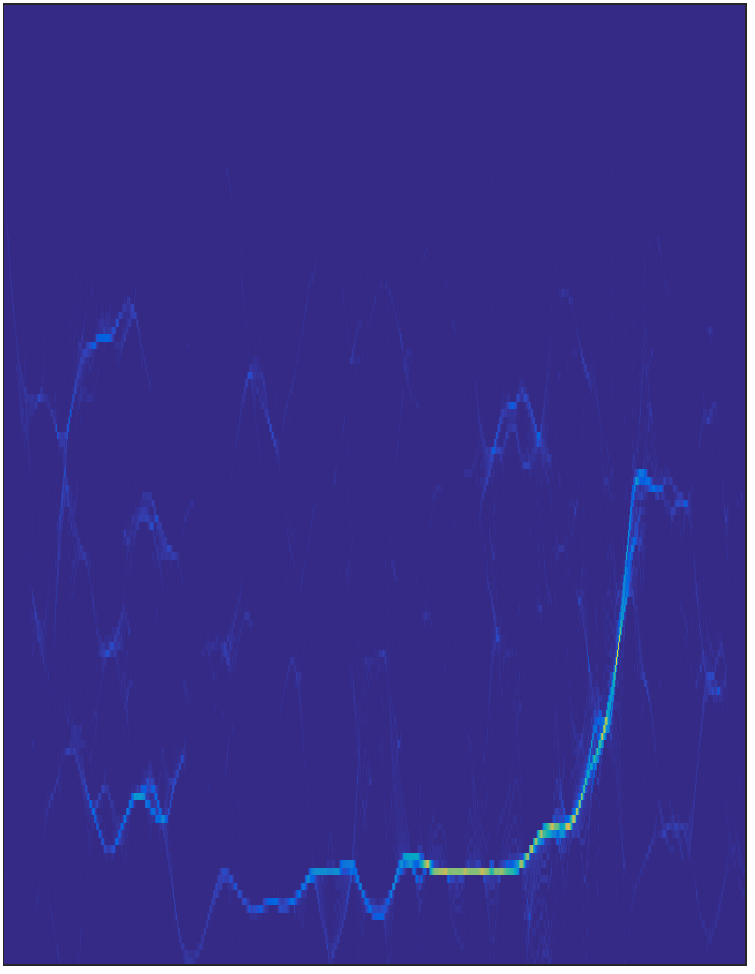}}
\centerline{(d)} 
\end{minipage}
\caption{Illustration of the TF representations of the gravitational-wave event \textbf{GW150914}, (a): observed Hanford signal; (b): STFT; (c): FSST2; (d): FSST4.}
\label{fig:Fig7}
\end{figure*}

\begin{figure}[!htb]
\begin{minipage}[b]{0.48\linewidth}
\centering
\centerline{\includegraphics[width = 0.75\linewidth]{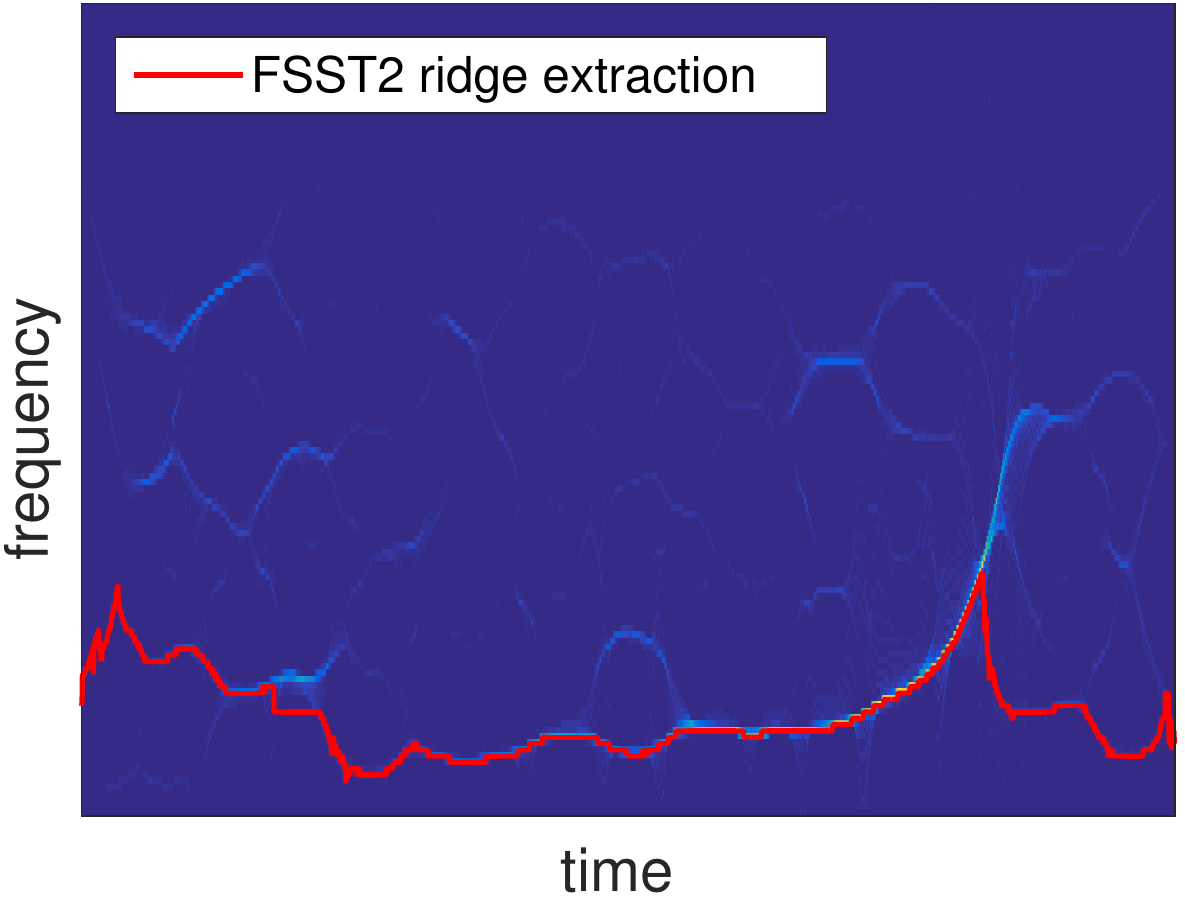}}
\centerline{(a)} 
\end{minipage}
\begin{minipage}[b]{0.48\linewidth}
\centering
\centerline{\includegraphics[width = 0.75\linewidth]{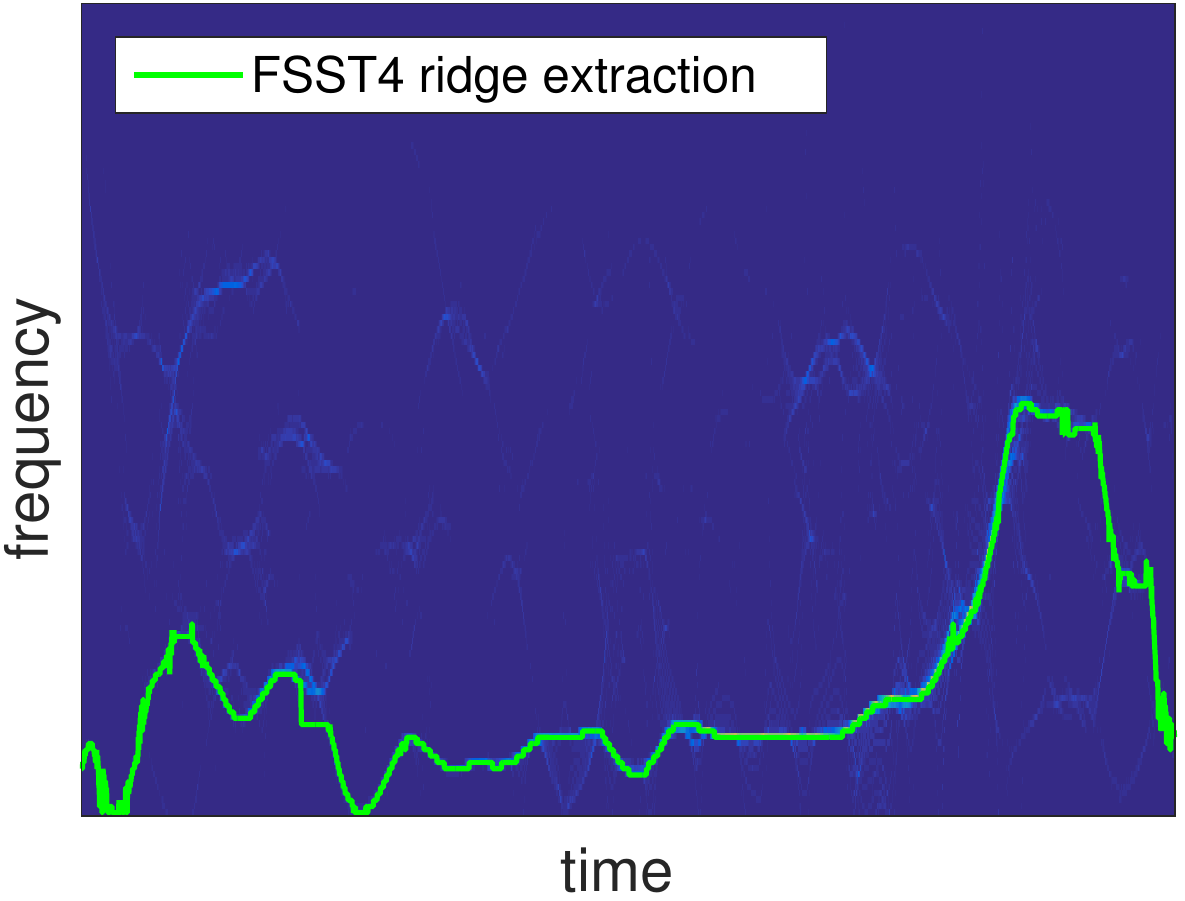}}
\centerline{(b)} 
\end{minipage}\\
\begin{minipage}[b]{0.48\linewidth}
\centering
\centerline{\includegraphics[width = 0.75\linewidth]{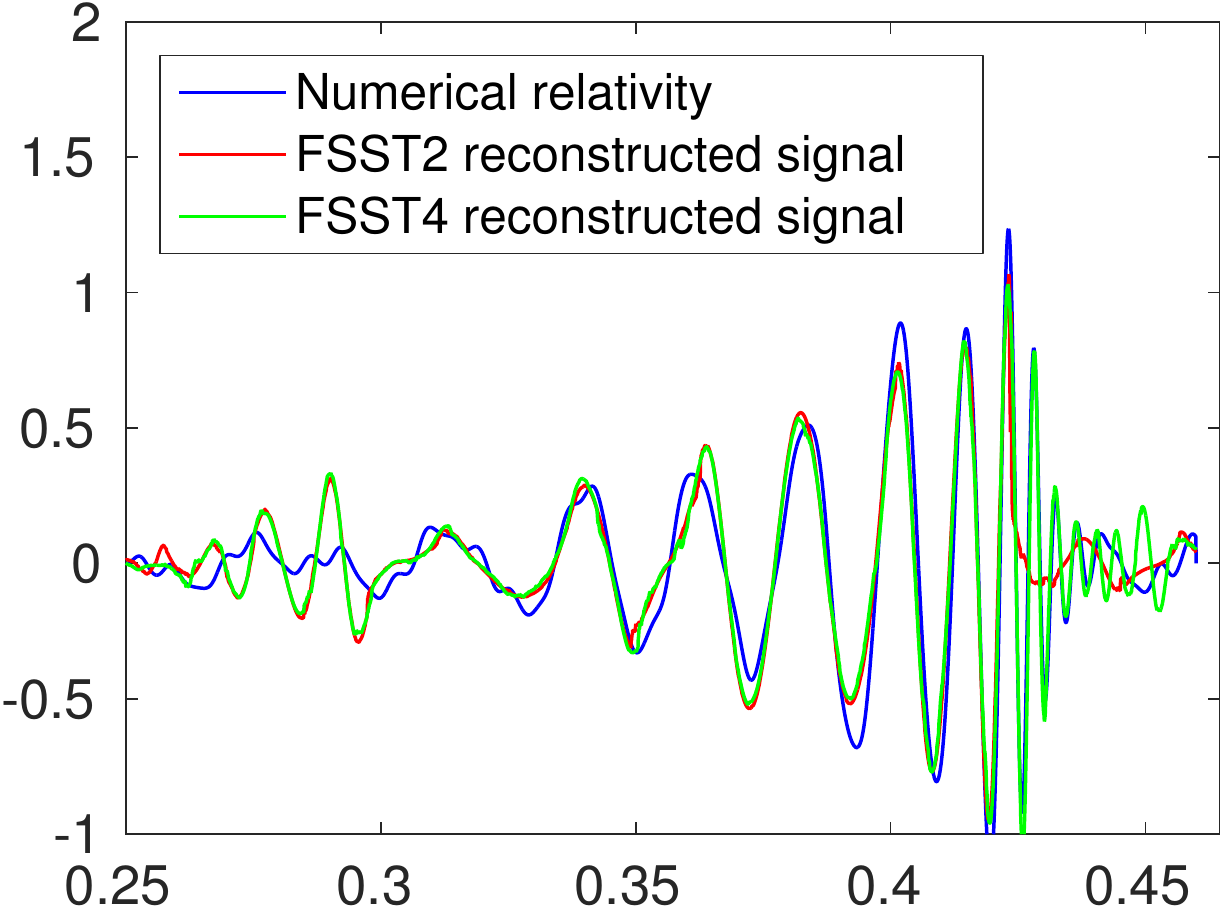}}
\centerline{(c)} 
\end{minipage}
\begin{minipage}[b]{0.48\linewidth}
\centering
\centerline{\includegraphics[width = 0.75\linewidth]{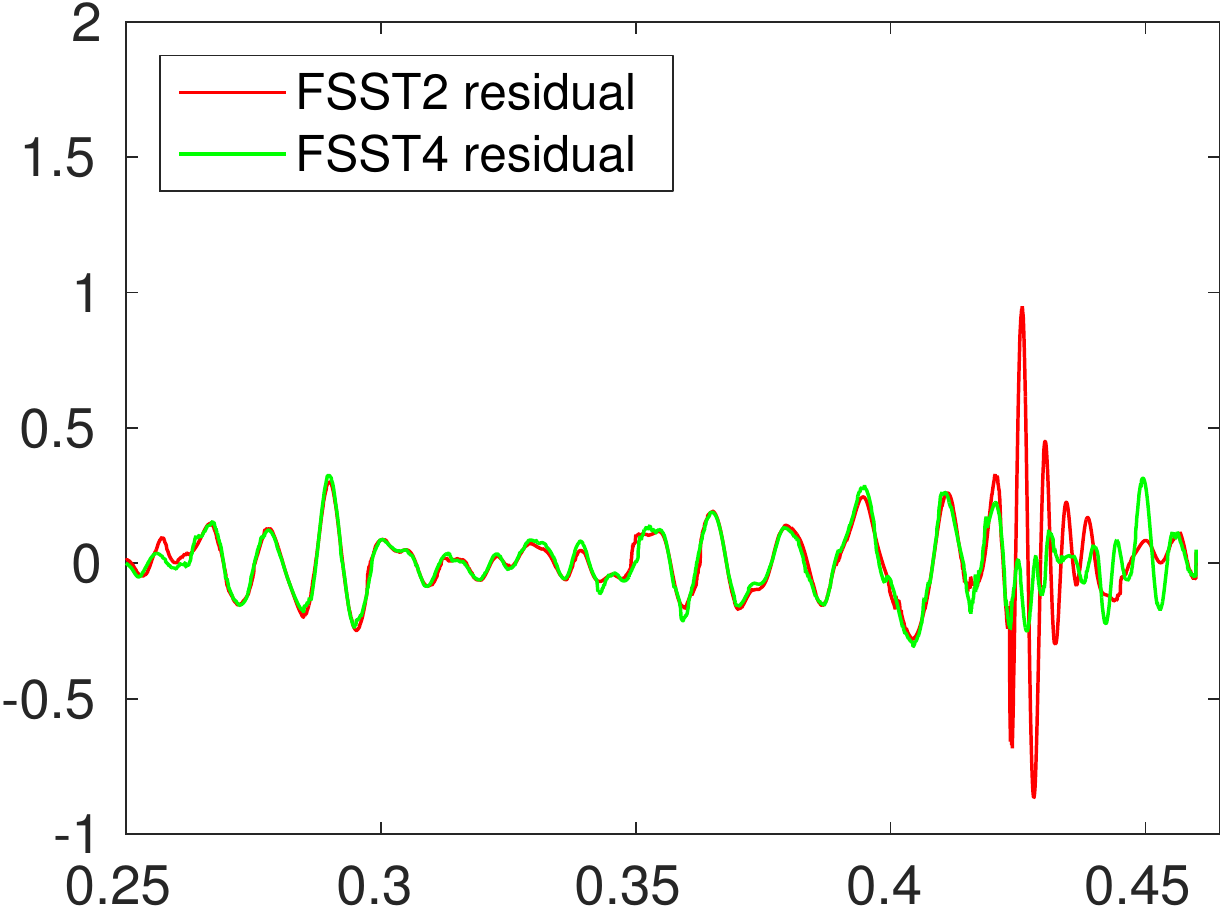}}
\centerline{(d)} 
\end{minipage}
\caption{(a): the ridge estimated from FSST2 displayed in Figure \ref{fig:Fig7}  (c); (b): same as (a) but on FSST4; (c): reconstructed signals along with numerical relativity waveform for a system with parameters consistent with those recovered from \textbf{GW150914} event confirmed by an independent computation; (d): their corresponding residuals after subtracting the numerical relativity waveform.}
\label{fig:fig:Fig8}
\end{figure}

We first display the gravitational-wave strain observed by the LIGO Hanford in Figure \ref{fig:Fig7} (a), and the STFT, the reassigned transforms corresponding 
to FSST2 and FSST4 in Figure \ref{fig:Fig7} (b), (c) and (d), respectively. 
The sharpened representations provided by FSST2 and FSST4 make the TF information more easily interpretable: as a matter of fact, the gravitational-wave signal consists of only one mode sweeping sharply upwards. However, the improvement brought by using high-order synchrosqueezing transform is not obvious at this point. 
 Moving on to mode reconstruction,  we perform ridge detection on each of the TF representations given by FSST2 and FSST4 and display the 
 results in Figure \ref{fig:fig:Fig8} (a) and (b).  We remark that FSST4 enables a better ridge detection of the three stages of the collision
 of two black-holes; especially the ``ring down'' one, which commences when the IF of the mode starts to decrease. This is associated with a sudden variation of the curvature 
 of its IF which is better taken into account by FSST4. A consequence of this can be seen in  Figure \ref{fig:fig:Fig8} (c) displaying 
 the reconstructed mode using either  FSST2 or FSST4, the latter leading to a much better reconstruction, very similar to the numerical relativity waveform 
 obtained from an independent calculation \cite{Abbott2016}. This fact is finally reflected by Figure \ref{fig:fig:Fig8} (d) in which we display 
 the residual errors (in $l_2$ norm) between the mode predicted by the numerical relativity  and the one reconstructed from  FSST2 or FSST4. 
 This thus demonstrates the interest of the proposed new technique in real applications. 

\section{Conclusion} \label{sec:conclusion}

In this paper, we introduced a generalization of the short-time Fourier-based synchrosqueezing transform by defining new synchrosqueezing operators based on high order amplitude and phase approximations. Such a generalization allows us to better handle a wide variety of multicomponent signals containing very strongly modulated AM-FM modes. The interest of the proposed new technique was also demonstrated through numerical experiments both for simulated and real signals. Indeed, it successfully produces a TF picture more concentrated than other methods based on synchrosqueezing or reassignment, while allowing for a better invertibility of the TF representation. 
Future work should now be devoted to the theoretical analysis of the behavior of the proposed representations when applied to noisy signals, as was done in \cite{Thakur2013,Yang2016} for the original FSST. In this regard, it would also be of interest to study  the behavior of the transform when the type of noise is 
non Gaussian.                               

\appendices
\section{The proof of Proposition \ref{propo:estimates_pj}} \label{app:estimates_pj}
\begin{IEEEproof}
First of all, we rewrite the expression (\ref{eqn:complex_IF_nSST_general}) under matrix form: 
 \begin{align} 
 \tilde{\omega}_{f} (t,\eta) &= \mathbf{X}_{N}(t,\eta)\cdot \mathbf{R}_{N}(t)^T \notag
\end{align} 
where $\mathbf{Z}^T$ is  the transpose of matrix $\mathbf{Z}$ and the two row vectors $\mathbf{X}, \mathbf{R}$ defined as: 
\begin{align}
 \mathbf{X}_{N}(t,\eta)& = \left[1~~x_{2,1}(t,\eta)~~... ~~x_{N,1}(t,\eta)\right]\notag\\
 \mathbf{R}_{N}(t)& = \left[r_{1}(t)~~ r_{2}(t)~~ ... ~~r_{N}(t)\right] \notag
\end{align}
Let us denote $y_{1}   =  \mathbf{X}_{N} \cdot \mathbf{R}_{N}^T$, we may thus write:
\begin{align} \label{eqn:first_equation_MX}
y_{1} 
=
\begin{bmatrix}
    x_{1,1} & x_{2,1} & x_{3,1} & \dots  & x_{N,1}  
\end{bmatrix}\mathbf{R}_{N}^T .
\end{align}
It is noteworthy that $\Re \left\{r_{1}(t) \right\}= \phi'(t)$. To get $r_{k}$, we build up a system of N equations with variables $r_{k}$ for $k = 1, \hdots, N$ from (\ref{eqn:first_equation_MX}) using the following procedure. By computing the partial derivatives of (\ref{eqn:first_equation_MX}) with respect to $\eta$ and using notation $y_{2}  = \dfrac{\partial_{\eta} y_{1} }{\partial_{\eta}x_{2,1}}$ and $x_{k,2} = \dfrac{\partial_{\eta}x_{k,1}}{\partial_{\eta}x_{2,1}}$, the second equation can be obtained:
\begin{align} 
y_{2} 
=
\begin{bmatrix}
    0 & 1 & x_{3,2} & \dots  & x_{N,2}  
\end{bmatrix}\mathbf{R}_{N}^T. \notag
\end{align}
Doing the same thing iteratively, we can get the $jth$ equation:
\begin{align} 
y_{j} 
=
\begin{bmatrix}
    0 & 0 & \dots & 1 & \dots  & x_{N,j}  
\end{bmatrix}\mathbf{R}_{N}^T. \notag
\end{align}
Combining all these equations, the desired system of equations can be deduced:
\begin{align} 
\begin{bmatrix}
   y_{1}  \\
   y_{2}  \\
   \vdots \\
   y_{N-1}  \\
   y_{N} 
 \end{bmatrix}
=
\begin{bmatrix}
    1 & x_{2,1} & x_{3,1} & \dots  & x_{N,1}  \\
    0 & 1 & x_{3,2}  &\dots  & x_{N,2} \\
    \vdots & \vdots  & \ddots & \vdots & \vdots \\
    0 & 0 & 0  & \dots  & x_{N,N-1} \\
    0 & 0 & 0  & \dots  & 1 
\end{bmatrix}\begin{bmatrix}
   r_{1} \\
   r_{2} \\
   \vdots \\
   r_{N-1} \\
   r_{N} \end{bmatrix} \notag
\end{align}
or 
\begin{align} \label{eqn:upper_matrix}
 \begin{bmatrix} \mathbf{Y}_{N} \end{bmatrix} = \begin{bmatrix} \mathbf{MX}\end{bmatrix}   \begin{bmatrix}\mathbf{R}_{N} \end{bmatrix}^T.
\end{align}
Since $\mathbf{MX}$ is a upper triangular matrix with nonzero diagonal coefficients, we use back-substitution algorithm to get $r_{k}$ for $k = 1, \hdots, N $ as follows: 

\begin{align} 
&r_{N}(t) = y_{N}(t,\eta)~\text{and} \notag\\
&r_{j}(t) = y_{j}(t,\eta) - \sum \limits_{k=j+1}^N x_{k,j}(t,\eta) r_{k}(t)~~~\text{for}~j = N-1, N-2, \hdots, 1. \notag
\end{align}
As a result, $\tilde{q}_{\eta,f}^{[k,N]}(t,\eta) = r_k(t)$ for $k = 2, \hdots, N$. From (\ref{eqn:complex_IF_nSST_general}), we clearly have: $\Re \left\{ \tilde{q}_{\eta,f}^{[k,N]}(t,\eta) \right\} = \dfrac{\phi^{(k)}(t)}{(k-1)!}$ for $k = 2, \hdots, N$, which finishes the proof. 
\end{IEEEproof}

\section{Proof of the Proposition \ref{propo:operators_efficient}} \label{app:operators_efficient}
\begin{IEEEproof}
By using $\partial_{\eta}V_{f}^{t^{k-1}g} = -i2\pi V_{f}^{t^{k}g} ~\mathrm{for}~ k \in  \mathbb{N}$ and defining $X_{k,j} = V_{f}^{g}V_{f}^{t^{k}g}-V_{f}^{t^{j-1}g}V_{f}^{t^{k-j+1}g}$, we get the following formula: 
\begin{align}
\partial_{\eta}X_{k,j} &= X_{k+1,j}  + X_{k+1,j} - X_{k+1,2}. \notag
 \end{align}
Thus, the upper triangular part of matrix $\mathbf{MX}$ defined in (\ref{eqn:upper_matrix}) with $N=4$ can be obtained as follows:  
\begin{align}
 &x_{k,1} = \frac{V_{f}^{t^{k-1} g}}{V_{f}^{g}}~~~~~~~~~~~\mathrm{~~~~~for}~ k=1 \hdots 4, \notag\\
 &x_{k,2} = \dfrac{\partial_{\eta}x_{k,1}}{\partial_{\eta}x_{2,1}} = \dfrac{V_{f}^{g} V_{f}^{t^{k}g}-V_{f}^{tg} V_{f}^{t^{k-1}g}}{V_{f}^{g} V_{f}^{t^{2}g}-\left(V_{f}^{tg}\right)^2}  = \dfrac{X_{k,2}}{X_{2,2}}~\mathrm{~~~~~for}~ k=2 \hdots 4, \notag\\
 & x_{k,3} = \dfrac{\partial_{\eta}x_{k,2}}{\partial_{\eta}x_{3,2}} = \dfrac{X_{k+1,3}X_{2,2} - X_{k,2} X_{3,3}}{X_{4,3}X_{2,2} - X_{3,2}X_{3,3}}~~ \mathrm{~~~~~~~~~~~for}~ k=3 \hdots 4, \notag\\
 &x_{k,4} = 1~~~~~~~~~~~\mathrm{~~~~~for}~ k = 4  \notag
 \end{align}

Also, the elements of vector $\mathbf{Y}$ are obtained by:    
\begin{align} 
y_{1}  &= \tilde{\omega}_{f} =\eta - \frac{1}{i2\pi} \frac{V_{f}^{g'}}{V_{f}^g},\notag\\
y_{2}  &=  \dfrac{\partial_{\eta} y_{1} }{\partial_{\eta}x_{2,1}} \notag \\
&= \frac{1}{i2\pi}\frac{\left(V_{f}^{g}\right)^2+V_{f}^{g}V_{f}^{tg'}-V_{f}^{tg}V_{f}^{g'}}{V_{f}^{g} V_{f}^{t^{2}g}-\left(V_{f}^{tg}\right)^2} = \dfrac{W_{2}}{X_{2,2}}, ~\mathrm{with}~~  W_{2} = \frac{1}{i2\pi} \left[\left(V_{f}^{g}\right)^2+V_{f}^{g}V_{f}^{tg'}-V_{f}^{tg}V_{f}^{g'}\right]. \notag\\ 
y_{3}  & = \dfrac{\partial_{\eta} y_{2}}{\partial_{\eta}x_{3,2}}= \dfrac{W_{3}X_{2,2} - W_{2} X_{3,3}}{X_{4,3}X_{2,2} - X_{3,2}X_{3,3}},~\mathrm{with}~~ W_{3} = \partial_{\eta} W_{2}. \notag\\
y_{4}  &= \dfrac{\partial_{\eta} y_{3}}{\partial_{\eta}x_{4,4}} =\dfrac{
\splitdfrac{\splitdfrac{\left(X_{4,3}X_{2,2} - X_{3,2}X_{3,3}\right)W_{4}}{-\left(W_{3}X_{2,2} - W_{2} X_{3,3}\right)\left(X_{5,4}  + X_{5,3} - X_{5,2}\right)}}{
+\left(W_{3}X_{3,2}-  W_{2} X_{4,3} \right) \left( X_{4,4} + X_{4,3} - X_{4,2}\right)}}{\splitdfrac{\splitdfrac{\left(X_{4,3}X_{2,2} - X_{3,2}X_{3,3} \right)\left(X_{6,4} + X_{6,3} - X_{6,2} \right)}{{}-\left(X_{5,3}X_{2,2} - X_{4,2} X_{3,3}\right)\left(X_{5,4} + X_{5,3} - X_{5,2}\right)}}{
+\left(X_{5,3} X_{3,2} - X_{4,2} X_{4,3}\right)  \left( X_{4,4} + X_{4,3} - X_{4,2} \right)}}, ~ \mathrm{with}~~ W_{4} = \partial_{\eta} W_{3}.  \notag
\end{align}
With the help of back-substitution algorithm, the modulation operators read: 
\begin{align}
&\tilde{q}_{\eta,f}^{[4,4]} = \dfrac{
\splitdfrac{\splitdfrac{\left(X_{4,3}X_{2,2} - X_{3,2}X_{3,3}\right)W_{4}}{-\left(W_{3}X_{2,2} - W_{2} X_{3,3}\right)\left(X_{5,4}  + X_{5,3} - X_{5,2}\right)}}{
+\left(W_{3}X_{3,2}-  W_{2} X_{4,3} \right) \left( X_{4,4} + X_{4,3} - X_{4,2}\right)}}{\splitdfrac{\splitdfrac{\left(X_{4,3}X_{2,2} - X_{3,2}X_{3,3} \right)\left(X_{6,4} + X_{6,3} - X_{6,2} \right)}{{}-\left(X_{5,3}X_{2,2} - X_{4,2} X_{3,3}\right)\left(X_{5,4} + X_{5,3} - X_{5,2}\right)}}{
+\left(X_{5,3} X_{3,2} - X_{4,2} X_{4,3}\right)  \left( X_{4,4} + X_{4,3} - X_{4,2} \right)}} \notag \\[0.5em]
&\tilde{q}_{\eta,f}^{[3,4]} = \dfrac{W_{3}X_{2,2} - W_{2} X_{3,3}}{X_{4,3}X_{2,2} - X_{3,2}X_{3,3}} - \tilde{q}_{\eta,f}^{[4,4]}\dfrac{X_{5,3}X_{2,2} - X_{4,2} X_{3,3}}{X_{4,3}X_{2,2} - X_{3,2}X_{3,3}} \notag\\[0.5em]
&\tilde{q}_{\eta,f}^{[2,4]} = \dfrac{W_{2}}{X_{2,2}} - \tilde{q}_{\eta,f}^{[3,4]} \dfrac{X_{3,2}}{X_{2,2}} - \tilde{q}_{\eta,f}^{[4,4]}\dfrac{X_{4,2}}{X_{2,2}}. \notag
\end{align}

Finally, we complete the proof of this proposition by using notation $G_{k}$ and $G_{j,k}$ to rewrite the above expressions.
\end{IEEEproof} 
\bibliographystyle{IEEEtran}
\bibliography{FINALVERSION}
\end{document}